\documentclass[aps,prb,showpacs,preprintnumbers,twocolumn,superscriptaddress]{revtex4-2}
\usepackage{amsmath,amssymb}
\usepackage{bm}
\usepackage{tipa}
\usepackage{upgreek}
\usepackage{tabularx}
\usepackage{graphicx}
\usepackage{amsmath}
\usepackage{comment}
\usepackage{mathrsfs}
\usepackage{graphicx}
\usepackage{braket}
\usepackage{enumitem}
\usepackage{natbib}
\usepackage{mathbbol}
\usepackage{mathtools}
\usepackage{booktabs}
\usepackage{gensymb}
\usepackage{multirow}
\usepackage[normalem]{ulem}
\usepackage{color}
\usepackage[colorlinks,bookmarks=true,citecolor=blue,linkcolor=red,urlcolor=blue]{hyperref}
\usepackage{hyperref}
\renewcommand{\vec}[1]{\mathbf{#1}}
\usepackage{pifont}

\usepackage[caption=false]{subfig}
\usepackage{ragged2e} 
\DeclareCaptionJustification{justified}{\justifying}

\begin{document}

 \title{Phenomenology of bond and flux orders in kagome metals}

	\author{Glenn Wagner}
	\affiliation{Department of Physics, University of Zurich, Winterthurerstrasse 190, 8057 Zurich, Switzerland}

	\author{Chunyu Guo}
	\affiliation{Max Planck Institute for the Structure and Dynamics of Matter, Hamburg, Germany}

	\author{Philip J.~W.~Moll}
	\affiliation{Max Planck Institute for the Structure and Dynamics of Matter, Hamburg, Germany}

 	\author{Titus Neupert}
	\affiliation{Department of Physics, University of Zurich, Winterthurerstrasse 190, 8057 Zurich, Switzerland}
 
	\author{Mark H.~Fischer}
	\affiliation{Department of Physics, University of Zurich, Winterthurerstrasse 190, 8057 Zurich, Switzerland}

	\begin{abstract}
    Despite much experimental and theoretical work, the nature of the charge order in the kagome metals belonging to the family of materials AV$_3$Sb$_5$ (A=Cs,Rb,K) remains controversial.
    A crucial ingredient for the identification of the ordering in these materials is their response to external perturbations, such as strain or magnetic fields.
    To this end, we provide a comprehensive symmetry classification of the possible charge orders in kagome materials with a $2\times2$ increase of the unit cell.  Motivated by the experimental reports of time-reversal-symmetry breaking and rotational anisotropy, we consider the interdependence of flux and bond orders. Deriving the relevant Landau free energy for possible orders, we study the effect of symmetry-breaking perturbations such as strain and magnetic fields. Our results, thus, provide a roadmap for future tests of these intricate orders.
	\end{abstract}
	
 	\maketitle

\section{Introduction} 
Starting with the formation of crystalline materials, spontaneous symmetry breaking is a concept foundational to condensed matter physics and guides our categorical understanding of phases of matter \cite{SSB}. Charge order can break spatial symmetries in the form of nematicity (rotational symmetry) or density waves (translational symmetry), superconductors break particle number conservation, magnets break spin-rotation and time-reversal symmetry (TRS). Particularly interesting situations arise, when the order comprises multiple degenerate components, in other words when the irreducible representation (irrep) corresponding to the order is multi-dimensional or when order parameters from different irreps are (accidentally) almost degenerate. On the one hand, this situation can lead to additional breaking of symmetries:  in the case of a superconductor, a complex superposition of two order-parameter components leads to a chiral superconductor spontaneously breaking TRS \cite{Kallin_2016}. On the other hand, multi-component orders can also help restore symmetries: A $\vec{q} = (\pi,0)$ charge order breaks both translation and rotation symmetry, while a superposition of $(\pi,0)$ and $(0,\pi)$ charge order describes checkerboard order and restores rotational symmetry~\cite{checkerbrd}.

A charge order of unconventional nature has recently been identified in a family of quasi-two-dimensional metals AV$_3$Sb$_5$ (A=Cs,Rb,K), whose main structural motif is a kagome lattice of vanadium atoms \cite{Neupert,Nguyen}. The charge-ordered phase, which sets in at about 70--100~K, has been studied with a variety of experimental techniques including angle-resolved photoemission spectroscopy \cite{Ortiz2,Zhu}, scanning tunneling spectroscopy \cite{Zhao,Chen,Nie,Jiang,Hong}, nuclear magnetic resonance \cite{song2021orbital,Zheng,Nie}, X-ray scattering \cite{Chen2}, muon spin-relaxation measurements \cite{Mielke,yu2021evidence}, thermal \cite{zhao2021nodal} and electrical \cite{Ortiz,Du,Xiang,Nie,Guo,Huang,zhao2021nodal,guo2023correlated,guo2023distinct} transport, as well as magneto-optical Kerr effect \cite{hu2023timereversal,Xu,saykin2022high,farhang2023unconventional}.
Broad consensus exists that the charge order creates a $2\times2$ superstructure within the plane, whereas the out-of-plane ordering (such as $2\times2\times1$, $2\times2\times2$ or $2\times2\times4$) is still under debate \cite{conjoined,Christensen}. In addition, controversial results as to whether or not charge order spontaneously breaks time-reversal and/or rotational symmetry have been reported. Several phenomena usually associated with spontaneous TRS breaking (TRSB) can be induced by a (weak) magnetic field, such as a giant anomalous Hall effect~\cite{Yang,Yu2,Wang2} and non-vanishing Kerr rotations~\cite{saykin2022high}. However, whether the system breaks time-reversal symmetry spontaneously, in other words at zero magnetic field, has been challenged for instance by the absence of a Kerr effect in this regime~\cite{saykin2022high}. Furthermore, although there are some reports of TRSB at the charge-ordering temperature, there is a large increase in the respective signals around 30--50~K. Finally, while there are many reports of (rotational) anisotropy in these materials \cite{xiang2021twofold,Nie,Emergent,NLWang_twofold}, a recent study challenges these reports~\cite{guo2023correlated}.

The conflicting experimental reports of anisotropy and spontaneous TRSB naturally raise the question whether the experiments themselve change the state probed and if so, how the different ordering possibilities can still be distinguished experimentally. Importantly, the interplay of different components of the order parameter is crucial, since charge-density-wave order on the kagome lattice with an ordering vector $\vec{Q} = \vec{M}= \overline{\Gamma M}$ forms a three-dimensional irrep due to the three inequivalent $M$ points. A single order-parameter component, corresponding to a single $M$-point ordering vector, breaks the rotational symmetry, whereas an equal superposition of all three could restore the rotational symmetry of the lattice. Furthermore, a complex order, arising due to the superposition of (almost degenerate) bond order and flux order would lead to TRSB.

To understand the physics of charge density waves in the kagome systems better and discuss their coupling to external perturbations that could help identify them, such as magnetic fields or strain, the proper symmetry of the ordering possibilities need to be studied first. Such an analysis then allows for the derivation of an effective Landau description. 
While both electronic \cite{Park} and phonon \cite{Tan,Christensen} instabilities have been studied as mechanisms for the charge ordering in the literature, such an approach has the advantage of being agnostic to the microscopic mechanism of the ordering.

Here, we present a comprehensive symmetry classification of all in-plane charge orders---including onsite charge modulations, bond orders, and flux/orbital current orders---on the kagome lattice with $2\times 2$ unit cell following the scheme introduced in Ref.~\cite{Venderbos}. This classification scheme provides a transparent way to develop an effective theory in the spirit of a Landau free energy for all charge orders, their coupling to each other and to magnetic fields and strain. This theory is applicable to, but not limited to, AV$_3$Sb$_5$. Since flux and bond orders both renormalize electron hopping integrals, we consider them intertwined. However, with the two generically transforming as different irreps, we treat them as different, yet coupled, order parameters, instead of one single complex order parameter. We therefore do not adopt the usual presumption that one order parameter dominates at and near the phase transition, but we study the interplay of one flux and one bond order parameter. This consideration results in a few scenarios, which can be sharply distinguished experimentally through their behavior in a magnetic field and by the presence or absence of anisotropy. 

\section{Summary of results}

We start by considering only in-plane ordering. Motivated by experiments that have established a $2\times2$ in-plane increase in the unit cell size, we consider all possible charge, bond, and flux orders on the kagome lattice arising from nearest-neighbor interactions. In Sec.~\ref{sec:group_th}, we use a group theory analysis to classify all these orders within the framework introduced by Venderbos~\cite{Venderbos}. The translational-symmetry-breaking orders can be classified in four different irreps of the (enlarged) symmetry group, labelled $F_1,\dots,F_4$. The irreps are three-dimensional and $F_{1,2}$ are even under $C_2$ while $F_{3,4}$ are odd. Bond order $\vec\Delta$ can fall in any of the four translationally symmetry breaking irreps, while flux order $\vec\Delta'$ has to either fall in the $F_2$ or $F_4$ irrep. 

The observation of time-reversal-symmetry breaking shows that flux order is present (although whether or not a small magnetic field is necessary to establish the flux order is not clear). Bond and flux order naturally couple to one another and in Sec.~\ref{sec:GL}, we therefore consider Landau theories that include both orders. \textcolor{black}{The second and fourth order terms in the free energy are identical regardless of which orders are combined, however, the third-order terms can differ. If the bond order transforms under $F_1$, then a term of the form $\Delta^3$ is allowed. In addition, the bond and flux order can couple via $\Delta\Delta'^2$ (the flux order needs to appear squared for the free energy to respect time-reversal symmetry).  } 

To gain intuition about the Landau free energy, in Sec.~\ref{sec:pure_charge_order} we start by discussing the simpler case where only charge order is present. We discuss separately the cases where a third-order term is present or absent, since this significantly impacts the phase diagram. In Sec.~\ref{sec:coupled_charge_and_current_order}, we then discuss the more complicated case with both charge and flux order present. We again derive the phase diagrams for the cases with and without the third-order term. In the phase diagrams, we use the relative critical temperature of the bond and flux orders as a tuning parameter. We obtain three types of phases: Either only $\vec\Delta$ or only $\vec\Delta'$ is present, or both are present simultaneously. The latter two phases break time-reversal symmetry. These different phases may or may not spontaneously break the $C_6$ symmetry depending on the specific irreps of the respective orders. 

One way to distinguish the different possible order parameters is to consider the impact of symmetry-breaking perturbations. In Sec.~\ref{sec:symmetry_breaking}, we therefore investigate the effects of strain and an out-of-plane magnetic field. The lowest-order coupling to such a magnetic field $B$ takes the form $B\Delta\Delta'$. Since the magnetic field is odd under in-plane mirror symmetries, this term is only allowed when the product $\Delta\Delta'$ is odd under these mirrors.

Up to this point, the Landau theory considered was very general and relied only on well-established experimental facts on the kagome metals. In Sec.~\ref{sec:exp_constr}, we review the experimental situation in more detail and suggest that the most likely order parameter combination to describe the experiments is a $F_1$ bond order with a $F_2$ flux order. Note, however, that while this conclusion relies on experimental input that is less well-established, we emphasize that the general discussion does not. 

Finally, in Sec.~\ref{sec:exp_proposals} we propose several experiments including elastoresistance, STM and resonant ultrasound spectroscopy that in combination with the Landau analysis would allow to clearly establish the type of ordering in the kagome metals.

\section{Symmetry Analysis}
\label{sec:group_th}
In the following, we consider a single kagome layer with point group $C_{6v}$~\footnote{Note that the full three-dimensional point group is $D_{6h}=C_{6v}\otimes \sigma_h$ with $\sigma_h$ denoting the mirror $z\mapsto -z$, but we restrict our considerations to $C_{6v}$ for simplicity.}.
Further, we study translational symmetry breaking arising from $M$-point ordering vectors in the kagome Brillouin zone. This ordering vector arises due to three van-Hove singularities (VHS) of the band structure of the kagome lattice at the three $M$ points \cite{Wu}:\begin{equation}
\vec{M}_{1,3}=\frac{\pi}{a \sqrt{3}}( \pm \sqrt{3}, 1), \quad \vec{M}_2=\frac{2 \pi}{a \sqrt{3}}(0,-1),
\end{equation}
where $a$ is the lattice constant. Close to the van Hove filling, the low-energy physics is dominated by scattering between these VHS, with momentum transfers corresponding to momentum differences between the $M$-points. These nesting vectors are also $M$-point vectors, since $\vec{M}_2-\vec{M}_3 \equiv \vec{M}_1$ (up to a reciprocal lattice vector) and the order parameters consist of superpositions of waves with wavevectors $\vec M_i$. Therefore, the order parameter in the unit cell centered at $\vec{R}$ will be a linear superposition of the components of
\begin{equation}
\vec{v}(\vec{R})=\left(\begin{array}{c}
\cos \vec{M}_1 \cdot \vec{R} \\
\cos \vec{M}_2 \cdot \vec{R} \\
\cos \vec{M}_3 \cdot \vec{R}
\end{array}\right)
\end{equation}
leading to an increase in the size of the unit cell by $2\times2$. The corresponding Bragg peaks are indeed seen experimentally in X-ray diffraction \cite{Ortiz2,Ortiz3,conjoined} and STM \cite{Jiang,Zhao,Liang_STM,Chen}.

\begin{figure} [tt]
    \centering
    \includegraphics[width=0.7\columnwidth]{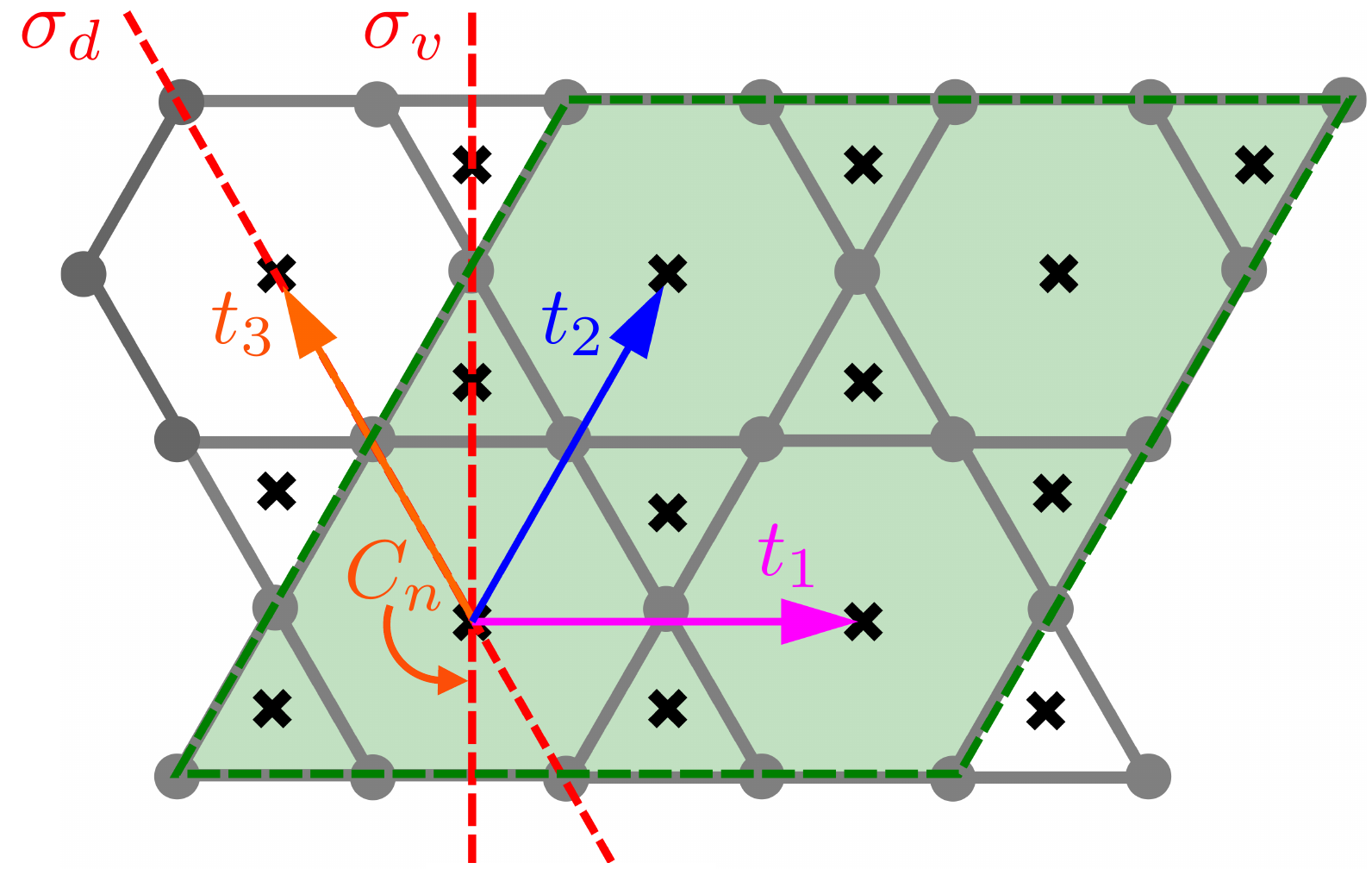}
    \caption{Schematic of the $2\times2$ enlarged unit cell of the kagome lattice with definitions of the elements of $C_{6v}'''$. We define the three translation operations $t_i$, mirror reflections $\sigma_v,\sigma_d$ and rotations $C_n$ with $n=2,3,6$. Sites and bonds are indicated in grey, while plaquettes are indicated by a black cross. There are 12 sites, 24 bonds and 12 plaquettes within the $2\times 2$ enlarged unit cell (green shading).}
    \label{fig:Kagome_symms}
\end{figure}

The possible bond and flux orders have been classified previously  in Refs.~\onlinecite{Feng} and \onlinecite{Christensen2} regarding the point group $D_{6h}$. We here choose a different route by following the classification scheme introduced in Ref.~\onlinecite{Venderbos} and restricting ourselves to the point group $C_{6v}$ for simplicity.
In this real-space scheme, the relevant symmetry group is enlarged to $C_{6v}'''$, where the primes indicate that the point group of the kagome lattice, $C_{6v}$, contains three additional elements corresponding to translations $\vec{t}_i$ (with $i=1,2,3$) that describe the translational symmetry breaking. The enlarged unit cell as well as the symmetry operations forming the group $C_{6v}'''$ are shown in Fig.~\ref{fig:Kagome_symms}. The group $C_{6v}'''$ has four one-dimensional and two two-dimensional irreps that are trivial under translation and are thus simply analogous to the irreps of $C_{6v}$. There are also four three-dimensional irreps $F_i$, which, in contrast, are non-trivial under translations. Their dimensionality directly follows from the fact that there are three $M$ points in the Brillouin zone. Table~\ref{tab:CT} presents the character table for the irreps of $C_{6v}'''$.

\begin{table}[t]
\begin{tabular}{c|c|c|c|c|c|c|c|c|c|c}
                               & $I$ & $t_i$ & $C_2$ & $t_iC_2$ & $C_3$ & $C_6$ & $\sigma_v$ & $t_i\sigma_v$ & $\sigma_d$ & $t_i\sigma_d$ \\ \hline
$|\mathcal{C}|$ & 1   & 3     & 1     & 3        & 8     & 8     & 6          & 6             & 6          & 6             \\ \hline
$A_1$                          & 1   & 1     & 1     & 1        & 1     & 1     & 1          & 1             & 1          & 1             \\
$A_2$                          & 1   & 1     & 1     & 1        & 1     & 1     & -1         & -1            & -1         & -1            \\
$B_1$                          & 1   & 1     & -1    & -1       & 1     & -1    & 1          & 1             & -1         & -1            \\
$B_2$                          & 1   & 1     & -1    & -1       & 1     & -1    & -1         & -1            & 1          & 1             \\
$E_1$                          & 2   & 2     & -2    & -2       & -1    & 1     & 0          & 0             & 0          & 0             \\
$E_2$                          & 2   & 2     & 2     & 2        & -1    & -1    & 0          & 0             & 0          & 0             \\\hline
$F_1$                          & 3   & -1    & 3     & -1       & 0     & 0     & 1          & -1            & 1          & -1            \\
$F_2$                          & 3   & -1    & 3     & -1       & 0     & 0     & -1         & 1             & -1         & 1             \\
$F_3$                          & 3   & -1    & -3    & 1        & 0     & 0     & 1          & -1            & -1         & 1             \\ 
$F_4$                          & 3   & -1    & -3    & 1        & 0     & 0     & -1         & 1             & 1          & -1           
\end{tabular}
\caption{Character table of the group $C_{6v}'''$ \cite{Venderbos}. The one- and two-dimensional irreps preserve the translation symmetry of the original kagome lattice, while the three-dimensional irreps lead to a $2\times2$ increase in the unit cell.}
\label{tab:CT}
\end{table}

There are three fundamental types of order, which, following the nomenclature adapted in Ref.~\onlinecite{Venderbos}, we denote as site, bond, and flux order, with the latter two being real and imaginary renormalizations of the hopping integrals. We deduce which irreps of $C_{6v}'''$ these orders on the kagome lattice decompose into, considering only nearest-neighbor order for the bond and flux order. To do so, we consider the permutation matrices $\mathcal{P}$ describing the action of the symmetry operators on the sites, bonds and fluxes. These permutation matrices are representations of $C_{6v}'''$ and, using the character Tab.~\ref{tab:CT}, they can be decomposed into irreps, see appendix \ref{app:decomp}.

Site order corresponds to a modulation of $\langle a^\dagger_{i\sigma}(\vec{R}) a_{i\sigma}(\vec{R})\rangle$, the local electron density, where $a^\dagger_{i\sigma}(\vec{R})$ creates an electron with spin $\sigma=\uparrow,\downarrow$ at a position $\mathbf{R}+\vec{\delta}_i$. Here, $\vec{R}$ points to the unit cell center and $\delta_i$ is the position of the sublattice site $i=A,B,C$ with respect to $\vec{R}$. Denoting the order-parameter components on each sublattice by a three-dimensional vector $\vec{s}_i$, the site order takes the form
\begin{equation}
     \langle a^\dagger_{i\sigma}(\vec{R}) a_{i\sigma}(\vec{R})\rangle=\vec{s}_i\cdot\vec{v}(\vec{R}).
\end{equation}
With 12 sites in the $2\times2$--increased unit cell, the representation is 12 dimensional. The possible site orders then decompose into the following irreps
\begin{equation}
    \mathcal{P}_s=A_1+E_2+F_1+F_3+F_4.
\end{equation}

\begin{figure}[t]
    \centering
    \includegraphics[width=\columnwidth]{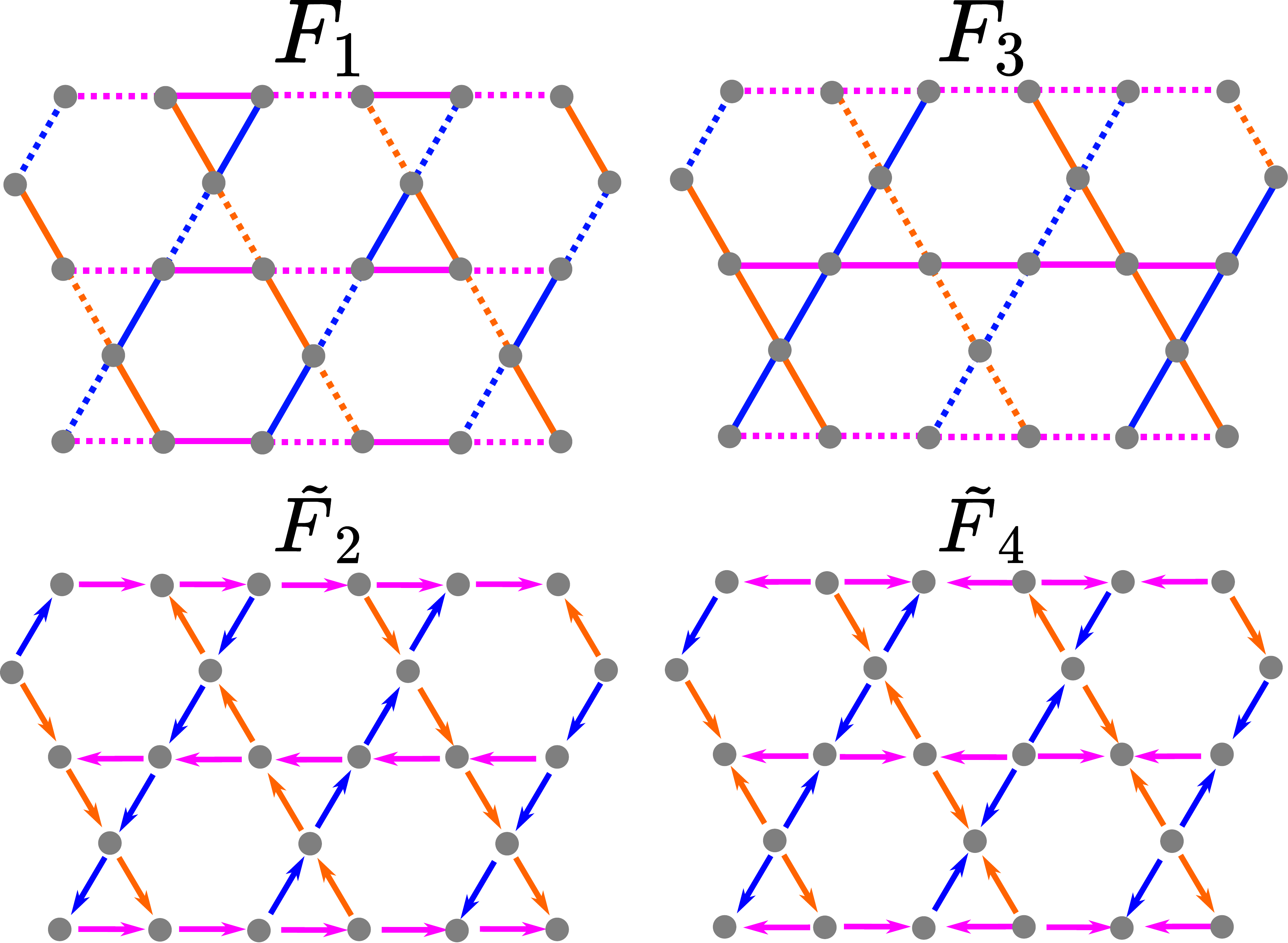}
    \caption{Schematics of the bond order $F_1, F_3$ and flux orders $ F_2',  F_4'$. For the bond orders, solid (dashed) lines indicate positive (negative) values of the order parameter. Colors indicate the components $\Delta_1,\Delta_2,\Delta_3$ of the bond order parameter. For the flux order parameter the arrows indicate the direction of the current.  Colors indicate the components $\Delta_1',\Delta_2',\Delta_3'$ of the flux order parameter.}
    \label{fig:KagomeOP}
\end{figure}

A general bond order corresponds to modulations
\begin{equation}
\begin{split}
    \langle a^\dagger_{A\sigma}(\vec{R}) a_{B\sigma}(\vec{R})\rangle&=\vec{w}_1\cdot\vec{v}(\vec{R})\\
    \langle a^\dagger_{A\sigma}(\vec{R}) a_{C\sigma}(\vec{R})\rangle&=\vec{w}_2\cdot\vec{v}(\vec{R})\\
    \langle a^\dagger_{A\sigma}(\vec{R}) a_{B\sigma}(\vec{R}-\vec{t}_3)\rangle&=\vec{w}_3\cdot\vec{v}(\vec{R})\\
    \langle a^\dagger_{A\sigma}(\vec{R}) a_{C\sigma}(\vec{R}+\vec{t}_2)\rangle&=\vec{w}_4\cdot\vec{v}(\vec{R})\\
    \langle a^\dagger_{B\sigma}(\vec{R}) a_{C\sigma}(\vec{R}+\vec{t}_2)\rangle&=\vec{w}_5\cdot\vec{v}(\vec{R})\\
    \langle a^\dagger_{C\sigma}(\vec{R}) a_{B\sigma}(\vec{R}-\vec{t}_3)\rangle&=\vec{w}_6\cdot\vec{v}(\vec{R}).
\end{split}
\end{equation}
We refer to real components of these $\vec{w}_i$ as bond order. There are 24 bonds within the $2\times 2$ unit cell and we find the bond order decomposition 
\begin{equation}
    \mathcal{P}_b=A_1+B_1+E_1+E_2+2F_1+F_2+2F_3+F_4.
\end{equation}
Finally, flux orders imply an imaginary component of the $\vec{w}_i$ \footnote{An exception would be specific gauges for flux orders with exactly $0$ or $\pi$ flux per plaquette. However, here we are interested in phases with generic values of flux.}. Equivalently, one can think of the resulting flux threading through the plaquettes. There are 12 plaquettes within the $2\times 2$ enlarged unit cell with possible orders decomposing into the irreps
\begin{equation}
    \mathcal{P}_\phi=2A_2'+ B_2'+2 F_2'+F_4'.
\end{equation}
In addition to spatial symmetries, the flux order breaks TRS, which we denote by a prime. Figure~\ref{fig:KagomeOP} shows examples of different types of translational symmetry breaking bond and flux orders for each of the four three-dimensional irreps. 

\begin{table*}[tt!]
\begin{tabularx}{\textwidth}{|l||c|c|c|c|X|X||X|X|X|X|X|}
\hline
      & $A_1$ & $A_2$ & $B_1$ & $B_2$ & $E_1$         & $E_2$         & $F_1$             & $F_2$             & $F_3$             & $F_4$             \\ \hline\hline
$A_1$ & $A_1$ & $A_2$ & $B_1$ & $B_2$ & $E_1$         & $E_2$         & $F_1$             & $F_2$             & $F_3$             & $F_4$             \\\hline
$A_2$ &       & $A_1$ & $B_2$ & $B_1$ & $E_1$         & $E_2$         & $F_2$             & $F_1$             & $F_4$             & $F_3$             \\\hline
$B_1$ &       &       & $A_1$ & $A_2$ & $E_2$         & $E_1$         & $F_3$             & $F_4$             & $F_1$             & $F_2$             \\\hline
$B_2$ &       &       &       & $A_1$ & $E_2$         & $E_1$         & $F_4$             & $F_3$             & $F_2$             & $F_1$             \\\hline
$E_1$ &       &       &       &       & $A_1+A_2+E_2$ & $B_1+B_2+E_1$ & $F_3+F_4$         & $F_3+F_4$         & $F_1+F_2$         & $F_1+F_2$         \\\hline
$E_2$ &       &       &       &       &               & $A_1+A_2+E_2$ & $F_1+F_2$         & $F_1+F_2$         & $F_3+F_4$         & $F_3+F_4$         \\\hline\hline
$F_1$ &       &       &       &       &               &               & $A_1+E_2+F_1+F_2$ & $A_2+E_2+F_1+F_2$ & $B_1+E_1+F_3+F_4$ & $B_2+E_1+F_3+F_4$ \\\hline
$F_2$ &       &       &       &       &               &               &                   & $A_1+E_2+F_1+F_2$ & $B_2+E_1+F_3+F_4$ & $B_1+E_1+F_3+F_4$  \\\hline
$F_3$ &       &       &       &       &               &               &                   &                   & $A_1+E_2+F_1+F_2$ & $A_2+E_2+F_1+F_2$ \\\hline
$F_4$ &       &       &       &       &               &               &                   &                   &                   & $A_1+E_2+F_1+F_2$ \\\hline
\end{tabularx}
\caption{The product of two irreps $R_1$ and $R_2$ can be decomposed into irreps of $C_{6v}'''$. The table is symmetric and for conciseness we only populate the upper triangle of the matrix. }
\label{tab:MT}
\end{table*}

\section{Landau theory}
\label{sec:GL}

Having categorized the possible site, bond, and flux orders, we can construct the free energy within Landau theory for different order parameters and their combinations. Having constructing such free energies then allows us to map out possible phase diagrams of the kagome metals, which capture the interplay of these order parameters, before studying the effect of external perturbations. The different responses to external perturbations can provide distinguishing experimental signatures. 

Previous theoretical work has already studied several types of Landau free energies for the kagome metals. In particular, Ref.~\onlinecite{Christensen} studied coupling between an $M$-point and an $L$-point order, and Ref.~\onlinecite{YangTheory} studied the coupling between an imaginary $M$-point order and superconductivity. Finally, Refs.~\onlinecite{Denner,Grandi,Lin,Park,tazai2022chargeloop,Christensen2} studied coupling between real and imaginary $M$-point orders, which is the case we aim to study here. Such a combination of orders is motivated by the observation of TRSB in experiments, which indicates that flux order is present at least under certain circumstances. Furthermore, since flux and bond order are the imaginary and real components, respectively, of the same (nearest-neighbor) order parameter, it is natural to consider a theory including both. 

However, most of the previous literature considered the bond order parameter to be a complex number, hence mixing bond and flux order of our classification. While this might be physically motivated and suggests a proximity of one order to the other, the real and imaginary components generally transform as different irreps, which manifests itself in different critical temperatures in these combined theories. We, thus, follow a different approach and in the following use the results from the order-parameter classification of Sec.~\ref{sec:group_th} with the multiplication Tab.~\ref{tab:MT} to write a family of free energies for two coupled order parameters transforming under two (different) three-dimensional irreps: a time-reversal symmetric $F_i$ and a TRS breaking $ F_{j}'$. In particular, with the free energy transforming as a scalar, only combinations of irreps, whose decomposition includes $A_{1}$, can appear. 
For this purpose, we first derive the Landau free energy $\mathcal{F}[\vec{\Delta}, \vec{\Delta}']$ up to fourth order in the order parameters $\vec{\Delta}$ and $\vec{\Delta}'$, before studying their coupling to strain and (out-of-plane) magnetic fields.

\subsection{Homogeneous $M$-point free energy}
The quadratic terms in the free energy take the same form, irrespective of the irrep combination. We include the temperature dependence to the quadratic coefficients 
\begin{equation}
\begin{split}
    \mathcal{F}^{(2)}[\vec{\Delta}, \vec{\Delta}']=&\alpha(T-T_{\rm c})(\Delta_1^2+\Delta_2^2+\Delta_3^2)\\&+\alpha'(T-T_{\rm c}')(\Delta_1'^2+\Delta_2'^2+\Delta_3'^2).
\end{split}
\end{equation}
The parameter $T_{\rm c}$ ($T_{\rm c}'$) is the temperature, at which the coefficient for the quadratic term in $\vec{\Delta}$ ($\vec{\Delta}'$) changes sign. While this sign change signals that the solution with vanishing order parameter becomes unstable, this temperature does not necessarily coincide with the critical temperature at which the order parameter acquires a non-zero value. The third-order terms, as well as the coupling between the order parameters can shift the critical temperature away from $T_{\rm c}$ ($T_{\rm c}'$). 

\begin{table}[t]
\begin{tabular}{cc|cc|}
\cline{3-4}
                                                  &       & \multicolumn{2}{c|}{\textbf{Flux order}}                                           \\ \cline{3-4} 
                                                  &       & \multicolumn{1}{c|}{$ F_2'$}                             & $ F_4'$ \\ \hline
\multicolumn{1}{|c|}{\multirow{4}{*}{\textbf{Bond order}}} & $F_1$ & \multicolumn{1}{c|}{$F_1^3, F_1 F_2'^2, BF_1F_2'$ } &   $F_1^3, F_1F_4'^2$            \\ \cline{2-4} 
\multicolumn{1}{|c|}{}                            & $F_2$ & \multicolumn{1}{c|}{\textcolor{black}{none} }                                         &    \textcolor{black}{none}          \\ \cline{2-4} 
\multicolumn{1}{|c|}{}                            & $F_3$ & \multicolumn{1}{c|}{none }                                     &       $BF_3F_4'$        \\ \cline{2-4} 
\multicolumn{1}{|c|}{}                            & $F_4$ & \multicolumn{1}{c|}{none }                                     & none       \\ \hline
\end{tabular}
\caption{Table of the allowed third-order terms and couplings to the magnetic field for a free energy $\mathcal{F}_{ij}$ constructed from order parameters transforming under a time-reversal symmetric $F_i$ and a time-reversal symmetry breaking $F_j'$. \textcolor{black}{Consequently, the free energies $\mathcal{F}_{12}$ and $\mathcal{F}_{14}$ have third order terms.} The free energies $\mathcal{F}_{12}, \mathcal{F}_{34}$ have linear coupling to a magnetic field.} 
\label{tab:third_order}
\end{table}

\begin{table}
    \centering
    \begin{tabular}{c|c}
    symmetrized product of irreps & decomposition\\
    \hline
    $(F_{i} \otimes F_{i})^{\mathrm{S}}$ & $A_{1} \oplus E_{2} \oplus F_{1}$\\
    $(F_{1} \otimes F_{1} \otimes F_{1})^{\mathrm{S}}$ & $A_{1} \oplus 2F_{1} \oplus F_{2}$\\
    $(F_{2} \otimes F_{2} \otimes F_{2})^{\mathrm{S}}$ & $A_{2} \oplus F_{1} \oplus 2F_{2}$\\
    $(F_{3} \otimes F_{3} \otimes F_{3})^{\mathrm{S}}$ & $B_{1} \oplus 2F_{3} \oplus F_{4}$\\
    $(F_{4} \otimes F_{4} \otimes F_{4})^{\mathrm{S}}$ & $B_{2} \oplus F_{3} \oplus 2F_{4}$\\
    $(F_{i} \otimes F_{i} \otimes F_{i} \otimes F_{i})^{\mathrm{S}}$ & $2A_{1} \oplus 2E_{2} \oplus 2F_{1} \oplus F_{2}$.
    \end{tabular}
    \caption{Multiplication table for the symmetrized products of the $F_i$ irreps. }
    \label{tab:symmetrized_products}
\end{table}

\textcolor{black}{The third-order terms may or may not be allowed, depending on the transformation properties of the irreps.} In Tab.~\ref{tab:third_order}, we list the allowed third-order terms for the different order-parameter combinations. When the terms are allowed, they take the form~\cite{Denner,Grandi}
\begin{align}
\mathcal{F}^{(3,0)}[\vec{\Delta}, \vec{\Delta}']=&\beta_1\Delta_1\Delta_2\Delta_3\label{eq:F30}\\
\mathcal{F}^{(1,2)}[\vec{\Delta}, \vec{\Delta}']=&\beta_2(\Delta_1\Delta_2'\Delta_3'+\Delta_1'\Delta_2\Delta_3'+\Delta_1'\Delta_2'\Delta_3).\label{eq:F12}
\end{align}
\textcolor{black}{The term $\mathcal{F}^{(3,0)}$ is allowed if $A_1\subset [F_i\otimes F_i\otimes F_i]^\textrm{S}$, while the term $\mathcal{F}^{(1,2)}$ is allowed if $A_1\subset F_i\otimes [F_j'\otimes F_j']^\textrm{S}$. The superscript ``$\textrm{S}$" denotes the symmetrized product of irreps, see Tab.~\ref{tab:symmetrized_products}.}

Due to time-reversal symmetry, there is neither a linear nor a cubic term for $\vec{\Delta}'$. 
Note that the third-order term in Eq.~\eqref{eq:F12} couples $\vec{\Delta}$ and $\vec{\Delta}'$ in such a way that any finite $\vec{\Delta}'$ induces a finite $\vec{\Delta}$ order, but not the other way around. \textcolor{black}{In other words, while the TRS-preserving bond order may exist by itself, a finite $\vec{\Delta}'$ always induces bond order transforming as $F_1$.} In the following, we will see how the presence of these third-order terms significantly alters the phenomenology of the ordered phase as compared to the case without such terms.

The fourth-order terms are again generically the same for all order-parameter combinations~\cite{Denner,Grandi,Christensen2}
\begin{equation}
\begin{split}
    \mathcal{F}^{(4)}=
    &\lambda_1(\Delta_1^2+\Delta_2^2+\Delta_3^2)^2\\
    &+\lambda_2(\Delta_1'^2+\Delta_2'^2+\Delta_3'^2)^2\\
    &+\lambda_3(\Delta_1^2+\Delta_2^2+\Delta_3^2)(\Delta_1'^2+\Delta_2'^2+\Delta_3'^2)\\
    &+\lambda_4(\Delta_1^2\Delta_2^2+\Delta_1^2\Delta_3^2+\Delta_2^2\Delta_3^2)\\
    &+\lambda_5(\Delta_1'^2\Delta_2'^2+\Delta_1'^2\Delta_3'^2+\Delta_2'^2\Delta_3'^2)\\
&+\lambda_6(\Delta_1^2\Delta_2'^2+\Delta_1^2\Delta_3'^2+\Delta_2^2\Delta_3'^2\\&\quad+\Delta_1'^2\Delta_2^2+\Delta_1'^2\Delta_3^2+\Delta_2'^2\Delta_3^2)\\    &+\lambda_7(\Delta_1\Delta_1'\Delta_2\Delta_2'+\Delta_1\Delta_1'\Delta_3\Delta_3'+\Delta_2\Delta_2'\Delta_3\Delta_3').
\end{split}
\end{equation}
The inclusion of the fourth-order terms is necessary for the thermodynamic stability of the free energy. Furthermore, in the absence of third-order terms, the fourth-order terms determine the form of the symmetry-breaking combination, such as anisotropic or TRS-breaking, below $T_{\rm c}$. $\mathcal{F}^{(4)}$ can also couple $\vec{\Delta}$ and $\vec{\Delta}'$, such that, for example, the $\lambda_3$ term describes attraction (repulsion) between the order parameters for $\lambda_3<0$ ($\lambda_3>0$). Note, however, that due to the quadratic nature of this term, such an interaction between the order parameters only amounts to a change of the critical temperature of the secondary order parameter.

\subsection{Comment on three-dimensional ordering}
\label{sec:L_point}
So far, our discussion has been based on a purely two-dimensional model. Despite the layered structure of the kagome metals, it is possible that three-dimensionality is important. Since some experiments report a $2\times2\times2$ increase in the size of the unit cell \cite{conjoined} and DFT calculations report instabilities at the $L$-points \cite{Christensen}, we are driven to consider $L$-point charge ordering as well. There are again three inequivalent wavevectors, which in this case are 
\begin{equation}
\vec{L}_{1,3}=( \pm \frac{\pi}{a }, \frac{\pi}{a \sqrt{3}},\frac{\pi}{c}), \quad \vec{L}_2=(0,-\frac{2 \pi}{a \sqrt{3}},\frac{\pi}{c}),
\end{equation}
where $c$ is the lattice constant in the vertical direction. 

Here, pure $L$-point third-order terms in the free energy will be absent, since the three momenta do not add up to zero. However, the second- and fourth-order terms will be present and unchanged with respect to the previous case, since only even powers of the order parameters appear and hence the $L$-point momenta in the $z$-direction add up to zero. In this sense, the pure $L$-point \emph{charge} order has the same Landau theory as a pure $M$-point \emph{flux} order \cite{Christensen2}. In the case of the $M$-point order, it is TRS that forces the absence of a pure third-order term, whereas in the $L$-point case, it is $z$-momentum conservation. 

In general, if we combine an $M$-point charge order $\vec{\Delta}^M$ with $L$-point charge (or flux) order $\vec{\Delta}^L$ (or $\vec{\Delta}^{\prime L}$), then all the allowed third-order terms take the schematic form $\vec{\Delta}^M(\vec{\Delta}^L)^2$, $\vec{\Delta}^M(\vec{\Delta}^{\prime L})^2$. These terms are only present as long as the $M$-point order \textcolor{black}{belongs to $F_1$}. Therefore, three-dimensional $L$-point bond or flux order always induces a subsidiary $M$-point charge order.

Finally, there have also been reports of $2\times2\times4$ order \cite{FSMap}. This time there are six inequivalent wavevectors (which we call `quarter'-points/$Q$-points), which in this case are 
\begin{align}
\vec{Q}_{1,2}&=(  \frac{\pi}{a }, \frac{\pi}{a \sqrt{3}},\pm\frac{\pi}{2c}), \\\vec{Q}_{3,4}&=(0,-\frac{2 \pi}{a \sqrt{3}},\pm\frac{\pi}{2c}), \\\vec{Q}_{5,6}&=( - \frac{\pi}{a }, \frac{\pi}{a \sqrt{3}},\pm\frac{\pi}{2c}).
\end{align}
The second order term would be proportional to $(Q_1Q_2+Q_3Q_4+Q_5Q_6)$. Since there are six order parameters, there is a significant increase in the number of fourth order terms. Third order terms will be absent from a pure $\vec{Q}$-point theory due to $k_z$ momentum conservation. If we combine an $L$-point charge order $\vec{\Delta}^L$ with $Q$-point charge (or flux) order $\vec{\Delta}^Q$ (or $\vec{\Delta}^{\prime Q}$), then all the allowed third-order terms take the schematic form $\vec{\Delta}^L(\vec{\Delta}^Q)^2$, $\vec{\Delta}^L(\vec{\Delta}^{\prime Q})^2$. 

\section{Pure bond order}
\label{sec:pure_charge_order}
We start our discussion with pure bond order without any flux order. 
This  case has already been extensively covered in Ref.~\onlinecite{Christensen} and here we just summarize the most important results. The possible Landau theories only differ in the presence or absence of the third-order term, Eq.~\eqref{eq:F30}. \textcolor{black}{As can be seen from Tab.~\ref{tab:third_order}, $F_1$ order at the $M$ point haa a third-order term, while the $F_2$, $F_3$ and $F_4$ orders at the $M$ point, as well as any $L$-point orders lack these terms.} 

\subsection{Without third-order term}
\label{sec:Pure_charge_order_3absent}
In this case, we have a second-order phase transition, when the coefficient of the quadratic term switches sign, in other words exactly at $T_{\rm c}$. The specific form of the ordering is determined by the fourth-order terms. With only $\vec{\Delta}$ present, only the $\lambda_1$ and $\lambda_4$ fourth-order terms are present. $\lambda_1$ alone does not break any degeneracy, irrespective of its value. $\lambda_4>0$ then leads to an anisotropic solution immediately below the charge-ordering temperature: Only one of the components $\Delta_i$ will be non-zero. On the contrary, $\lambda_4<0$ favors an isotropic solution. There are two degenerate isotropic solutions, since the free energy is independent of the sign of the individual order-parameter components: For the $F_3$ and $F_4$ irreps, the cases with all components the same sign ($\Delta_1=\Delta_2=\Delta_3$) and one component with the opposite sign ($\Delta_1=-\Delta_2=-\Delta_3$ or cyclic variations) are degenerate and related by $C_2$.

\subsection{With third-order term}
In the presence of a third-order term in the free energy, the phase transition changes to first order. Generically, the free energy takes the form
\begin{equation}
    \mathcal{F}=\alpha (T-T_{\rm c}) \Delta^2+b\Delta^3+c\Delta^4,
\end{equation}
which undergoes a first-order transition at $\tilde{T}_{\rm c} = T_{\rm c} + b^2/(4\alpha c) > T_{\rm c}$, where the order parameter jumps to a finite value $\Delta_0=-\frac{b}{2c}$. 

Further, the third-order term lifts the degeneracy between the isotropic solutions. In particular, for $\lambda_4\leq0$ and $\beta_1<0$, the configuration with $\text{sign}(\Delta_1\Delta_2\Delta_3)>0$, referred to as tri-hexagonal ordering for the case of $F_1$, is favored. When $\beta_1>0$, the configuration $\text{sign}(\Delta_1\Delta_2\Delta_3)<0$, the so-called Star-of-David ordering for $F_1$, is favored.

Finally, we can consider the case where $\lambda_4>0$ in the presence of a third-order term. In that case, the first-order transition into the $|\Delta_1|=|\Delta_2|=|\Delta_3|$ order 
is followed by a crossover to $|\Delta_2|=|\Delta_3|\approx0<|\Delta_1|$ (or cyclic variations) at lower temperatures.

\section{Coupled bond and flux order}
\label{sec:coupled_charge_and_current_order}

We now consider the case, where bond order and flux order coexist. Following our analysis in Sec.~\ref{sec:group_th}, we only have to consider flux orders belonging to the $F_2'$ and $F_4'$ irreps, while for bond order, all three-dimensional irreps have to be considered.  Translated into our irrep classification, previous work has studied various combinations of bond and flux order parameters: In Ref.~\onlinecite{Denner}, the authors study the single order parameter combination $F_1$ and $F_4'$; Ref.~\onlinecite{Park} considers the combination $F_1$ and $F_2'$; Ref.~\onlinecite{Lin} considers $F_1$ and $F_4'$; and Refs.~\onlinecite{Grandi} and \onlinecite{Christensen2} consider a variety of different order-parameter combinations. 

We construct the free energy $\mathcal{F}_{i j}$ for coupling bond order $F_i$ with flux order $ F_j'$. In the absence of additional perturbations, there are two cases shown in Tab.~\ref{tab:third_order}: Firstly, we consider coupling bond order that \textcolor{black}{belongs to the $F_1$ irrep} to any flux order (from the $F_2'$ or $F_4'$ irrep). Then, the free energy 
\begin{equation}
    \mathcal{F}_{ij} 
   = \mathcal{F}^{(2)}+\mathcal{F}^{(3,0)}+\mathcal{F}^{(1,2)}+\mathcal{F}^{(4)}
\end{equation}
where \textcolor{black}{$(ij)=(12),(14)$} includes all third-order terms. Secondly, we consider coupling bond order that \textcolor{black}{belongs to the $F_2$, $F_3$ or $F_4$ irreps to any flux order (from the $F_2'$ or $F_4'$ irrep)}. The corresponding free energy
\begin{equation}
    \mathcal{F}_{ij}
   = \mathcal{F}^{(2)}+\mathcal{F}^{(4)}
\end{equation}
where \textcolor{black}{$(ij)=(22),(24),(32),(34),(42),(44)$}  has no third-order terms. For both of these two cases, we next present a phase diagram, where we tune the relative strength of the $\vec{\Delta}$ and $\vec{\Delta}'$ order by tuning their relative critical temperature $T_c-T_c'$. 

\begin{figure}
    \includegraphics[width=\columnwidth]{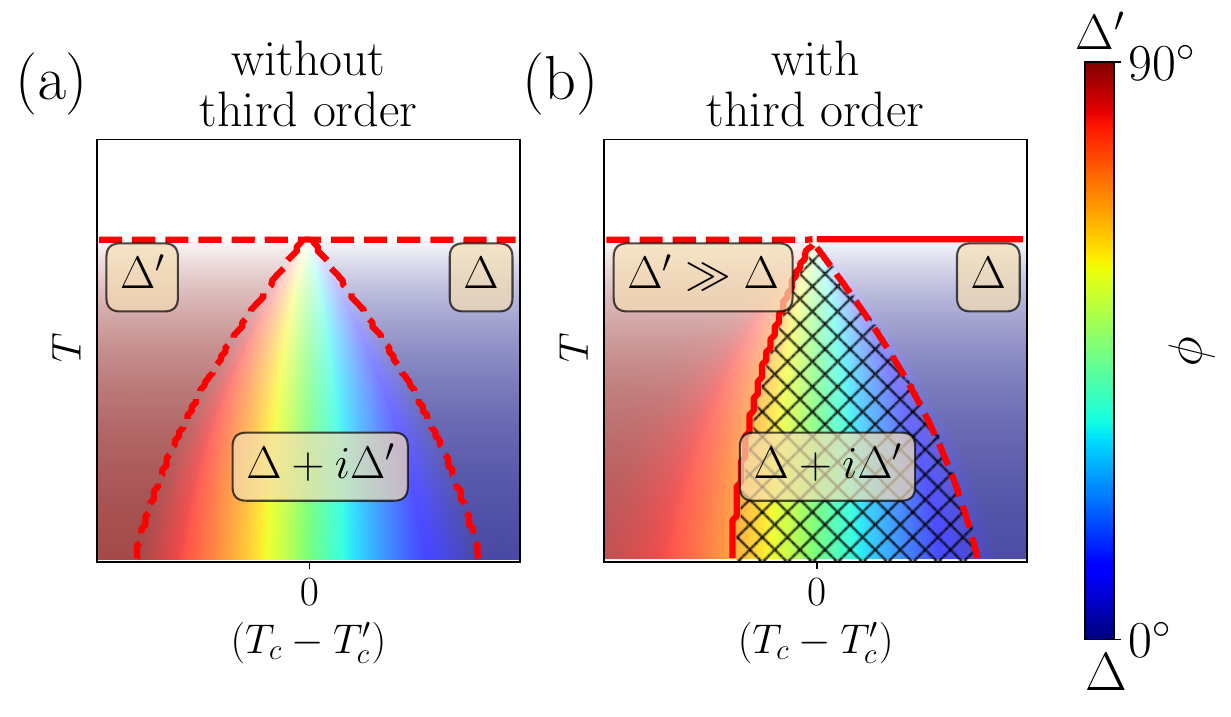}
    \caption{\textcolor{black}{(a) Phase diagram without a third order term applicable to the free energies $\mathcal{F}_{22}, \mathcal{F}_{24}, \mathcal{F}_{32}, \mathcal{F}_{34}, \mathcal{F}_{42}, \mathcal{F}_{44}$. (b) Phase diagram with a third order term applicable to the free energies $\mathcal{F}_{12}, \mathcal{F}_{14}$.} The color indicates the angle $\phi=\arctan \frac{\sum_i \Delta_i'^2}{\sum_i \Delta_i^2}$, while the intensity denotes $\sum_i (\Delta_i^2+\Delta_i'^2)$. Solid lines indicate first-order phase transitions and dashed lines indicate second-order transitions. Hatching denotes an anisotropic solution of the order parameter.  TRS is broken when $\Delta'>0$. The Landau free energy coefficients are chosen to be $\alpha=\alpha'=\lambda_1=\lambda_2=\mu_i=\beta_1=-\beta_2=1$, $\lambda_{i>2}=0$.}
    \label{fig:phase_diagrams_no_field}
\end{figure}

\subsection{Without third-order term}
Figure~\ref{fig:phase_diagrams_no_field}a shows the phase diagram without any third-order terms. This is the case for \textcolor{black}{$\mathcal{F}_{22}, \mathcal{F}_{24},\mathcal{F}_{32}, \mathcal{F}_{34}, \mathcal{F}_{42}, \mathcal{F}_{44}$}. Since $\vec{\Delta}$ and $\vec{\Delta}'$ are only coupled via the fourth-order term, we can have either order parameter existing alone. The phase diagram splits into three ordered regions that are entered through second-order transitions: one where $\vec{\Delta}$ exists alone ($T_c>T\gtrsim T_c'$), one where $\vec{\Delta}'$ exists alone ($T_c'>T\gtrsim T_c$) and one where both coexist. TRS is broken any time $\vec{\Delta}'$ is non-zero. Note that without a third-order term, there is a symmetry under exchanging $\vec{\Delta}$ and $\vec{\Delta}'$ (as well as exchanging the corresponding coefficients of the free energy). This explains the left-right symmetry in the phase diagram.

\subsection{With third-order term}
Figure~\ref{fig:phase_diagrams_no_field}b shows the phase diagram with third-order terms. This is the case for \textcolor{black}{$\mathcal{F}_{12}, \mathcal{F}_{14}$}. In this case, a finite $\vec{\Delta}'$ always induces a subsidiary $\vec{\Delta}$. Unlike the case without third-order terms, where there are three regions and the phase-transition into the coexistence region happens at a singular point ($T_{\rm c} = T_{\rm c}'$) and is always second order, here the phase diagram only has two regions: one, where $\vec{\Delta}$ appears alone and TRS is preserved and one, where both order parameters coexist and TRS is broken. As in the pure bond-order case, the transition into the former is always first order with a (potential) additional second-order transition into the coexistence region. The direct transition into the coexistence region changes from first order for $T_{\rm c} \approx T_{\rm c}'$ and $\Delta'\sim\Delta$ to second order for $T_{\rm c}' \gg T_{\rm c}$ and $\Delta'\gg\Delta$. Within the ordered phase, there is then a crossover between these domains. More details on the transitions can be found in appendix \ref{app:pd}.

Finally, the third-order term now breaks the symmetry of exchanging $\vec{\Delta}$ and $\vec{\Delta}'$, such that the phase diagram is no longer left-right symmetric.

\section{Symmetry-breaking}
\label{sec:symmetry_breaking}

\subsection{Coupling to strain}
Uniaxial strain has proven to be a very effective perturbation to probe correlated orders in two-dimensional and layered materials. Uniaxial strain was used in layered systems such as the cuprates to probe the charge density wave~\cite{choi:2022} or Sr$_2$RuO$_4$~\cite{grinenko:2021}, where it lifts the degeneracy of critical temperatures for the superconducting and TRS-breaking states, while the ground state in twisted bilayer graphene changes drastically under strain~\cite{nuckolls2023quantum}. Indeed, recent experiments indicate that coupling to strain significantly alters the transport properties of kagome metals~\cite{guo2023correlated} and we therefore include such coupling in our Landau theory. The strain matrix in terms of the displacement field $\mathbf{u}$ is given by
$\epsilon_{\alpha\beta}=\frac{1}{2}(\partial_\alpha u_\beta+\partial_\beta u_\alpha)$. The coupling to strain is independent of the order parameters we consider and has the form (see appendix \ref{app:strain})
\begin{equation}
\begin{split}
    \mathcal{F}^{(\textrm{str})}=&
    \mu_2[(\epsilon_{xx}-\epsilon_{yy})(\Delta_1^2-\Delta_2^2/2-\Delta_3^2/2)\\&\quad+\epsilon_{xy}\sqrt{3}(\Delta_2^2-\Delta_3^2)]\\
    &+\mu_3[(\epsilon_{xx}-\epsilon_{yy})(\Delta_1'^2-\Delta_2'^2/2-\Delta_3'^2/2)\\&\quad+\epsilon_{xy}\sqrt{3}(\Delta_2'^2-\Delta_3'^2)].
\end{split}
\end{equation}
Being quadratic in the order parameters, strain shifts the critical temperatures of different components of the order parameters and as such, it is possible that strain induces an order even though the temperature is too high in the strainless state. For example, it is possible for the strain to increase the critical temperature of one of the components of the flux order and thereby break TRS.

Finally, we note that having a first-order transition relies on the third-order term being relevant. The presence of strain favors some of the components of $\vec{\Delta}$ over others and, therefore, weakens the effect of the third-order term $\Delta_1\Delta_2\Delta_3$. Strain thus generally weakens the first-order nature of the transition into the ordered state. 

\subsection{Anisotropy}
\label{sec:nem}
As anisotropy we denote the breaking of rotation symmetry in a system. In the context of crystalline systems, where rotation symmetry is already discrete, anisotropy then refers to a further reduction, such as the breaking of $C_6$ down to $C_2$. In the context of a charge-density-wave instability with multiple inequivalent wave vectors, an anisotropy can arise from different magnitudes of the individual order-parameter components. Note that in the following, we consider purely in-plane order, since for finite out-of-plane momentum (such as $L$-point ordering), the rotational symmetry can be trivially broken \cite{Christensen}. 

Strain explicitly breaks rotation symmetry and will thus introduce an anisotropy. The susceptibility towards an anisotropy in a small strain field can thus serve as an indicator for the anisotropy in the system. In particular, the susceptibility of a $C_3$-breaking order parameter to strain will diverge for an anisotropic ground state. We thus calculate $\{\Delta_1^2-\Delta_2^2/2-\Delta_3^2/2,\frac{\sqrt{3}}{2}(\Delta_2^2-\Delta_3^2)\}$ to assess whether rotational symmetry is broken and, similarly, for the corresponding order parameter for flux order. In appendix \ref{app:strain}, we show how the susceptibility is related to this `order parameter' $\sum_{i,j}[(\Delta_i^2-\Delta_j^2)^2+(\Delta_i^{\prime2}-\Delta_j^{\prime2})^2]$. Below, we discuss the conditions for an anisotropy in the solution to the Landau free energy in the cases without and with third-order terms.

\subsubsection{Without third-order term}

As in the case of pure bond order, the fourth-order term can introduce an anisotropy. The terms with coefficients $\lambda_1$, $\lambda_2$ and $\lambda_3$ do not break the degeneracy between isotropic and anisotropic solutions. As shown in Sec.~\ref{sec:Pure_charge_order_3absent}, pure bond order is anisotropic when the fourth-order term $\lambda_4>0$. Similarly, flux order is anisotrpic when $\lambda_5>0$. In addition, $\lambda_6>0$ leads to an anisotropic solution, if both bond and flux order are non-zero. Note that the non-zero component of the two orders is the same, meaning a solution of the form $\Delta_i\neq0$, $\Delta_i'\neq0$ is stable. 

\subsubsection{With third-order term}

Anisotropy can arise from the third-order term in the free energy even when the fourth-order terms favor an isotropic solution . To see this, consider the terms
\begin{equation}
\beta_1\Delta_1\Delta_2\Delta_3+\beta_2(\Delta_1\Delta_2'\Delta_3'+\Delta_1'\Delta_2\Delta_3'+\Delta_1'\Delta_2'\Delta_3)
\end{equation}
as perturbations with $\Delta^2=\Delta_1^2+\Delta_2^2+\Delta_3^2$ and $\Delta'^2=\Delta_1'^2+\Delta_2'^2+\Delta_3'^2$ being fixed by the second order and fourth order terms in the free energy. Let us assume $\beta_1>0$ and $\beta_2<0$ (the case $\beta_1<0$ and $\beta_2>0$ can be obtained by flipping the sign of $\Delta_i$) . We can then compare the energies of two extreme cases: For an isotropic solution $\Delta_1=\Delta_2=\Delta_3=\Delta/\sqrt{3}$ and $\Delta_1'=\Delta_2'=\Delta_3'=\Delta'/\sqrt{3}$, the third order terms yield an energy
\begin{equation}
    \frac{|\beta_1|}{3\sqrt{3}}\Delta^3-\frac{|\beta_2|}{3\sqrt{3}}\Delta\Delta'^2
\end{equation}
while the anisotropic solution $\Delta_1=\Delta$ and $\Delta_2'=\Delta_3'=\Delta'/\sqrt{2}$ has energy
\begin{equation}
    -\frac{|\beta_2|}{2}\Delta\Delta'^2.
\end{equation}
Therefore the isotropic solution will be favoured when $(\Delta'/\Delta)^2>(\frac{3\sqrt{3}}{2}-1)\vert\beta_1\vert/\vert\beta_2\vert$. Increasing the ratio $\Delta'/\Delta$ corresponds to moving to the left in the phase diagram of Figure~\ref{fig:phase_diagrams_no_field}b. Similarly, the anisotropic solution gives way to an isotropic solution when $\Delta$ is large (see appendix \ref{sec:app_anis} for details). This leads to the wedge-shaped region in the phase diagram where the solution is anisotropic.

\subsection{Coupling to a z-axis magnetic field}
The lowest-order coupling to a magnetic field is linear in field and quadratic in the order parameters. However, only some of the order-parameter combinations couple to a magnetic field at this lowest order. The magnetic field breaks TRS and transforms under the $A_2$ irrep of $C_{6v}'''$. Therefore, for the order parameters to couple in this manner to the magnetic field, we require $A_2\subset F_i\otimes F_j'$.  If the coupling is allowed (see Tab.~\ref{tab:third_order}), it takes the form
\begin{equation}
    \mathcal{F}^{(B)}=\mu_1B(\Delta_1\Delta_1'+\Delta_2\Delta_2'+\Delta_3\Delta_3').
\end{equation}
We note that this is an unusual form of magnetic field coupling to (translational-symmetry-breaking) order parameters since the magnetic field couples \emph{linearly}. This is only possible since we have both TRS-breaking and TRS-preserving orders with the same wavevectors. Importantly, this term fixes the relative sign between $\vec{\Delta}$ and $\vec{\Delta}'$. Without a magnetic field, there can be domains with opposite relative signs. Applying a magnetic field causes the domains to flip such that the relative sign is the same everywhere. We note in passing that this form of the coupling to the magnetic field would also be possible if $\vec\Delta$ and $\vec\Delta'$ are both $L$-point orders, but not for a combination of an $L$-point and $M$-point order. 

\subsubsection{Without third-order term}
Figure~\ref{fig:phase diagrams}a shows the phase diagram in the presence of a magnetic field, when there is no third-order terms in the free energy. The only order-parameter combination that has the lowest-order coupling to a magnetic field, while lacking a third-order term, is $F_3$ with $F_4'$, in other words the free energy $\mathcal{F}_{34}$. In this case, the magnetic field couples $\vec{\Delta}$ and $\vec{\Delta}'$ and hence, the two orders always coexist. The regions $\Delta\gg\Delta'$, $\Delta\sim\Delta'$, and $\Delta\ll\Delta'$ are now separated by crossovers. In the region $T_c\sim T_c'$, the critical temperature is enhanced as both order parameters condense at the same time. While TRS is trivially broken in the entire phase diagram, the magnetic field does not induce an anisotropy if not present already.
Note that while in Ref.~\onlinecite{Tazai} the effect of magnetic field in a Landau theory of the kagome metals was considered, only  the case without third-order terms was treated.

\subsubsection{With third-order term}

\begin{figure}
    \centering
    \includegraphics[width=\columnwidth]{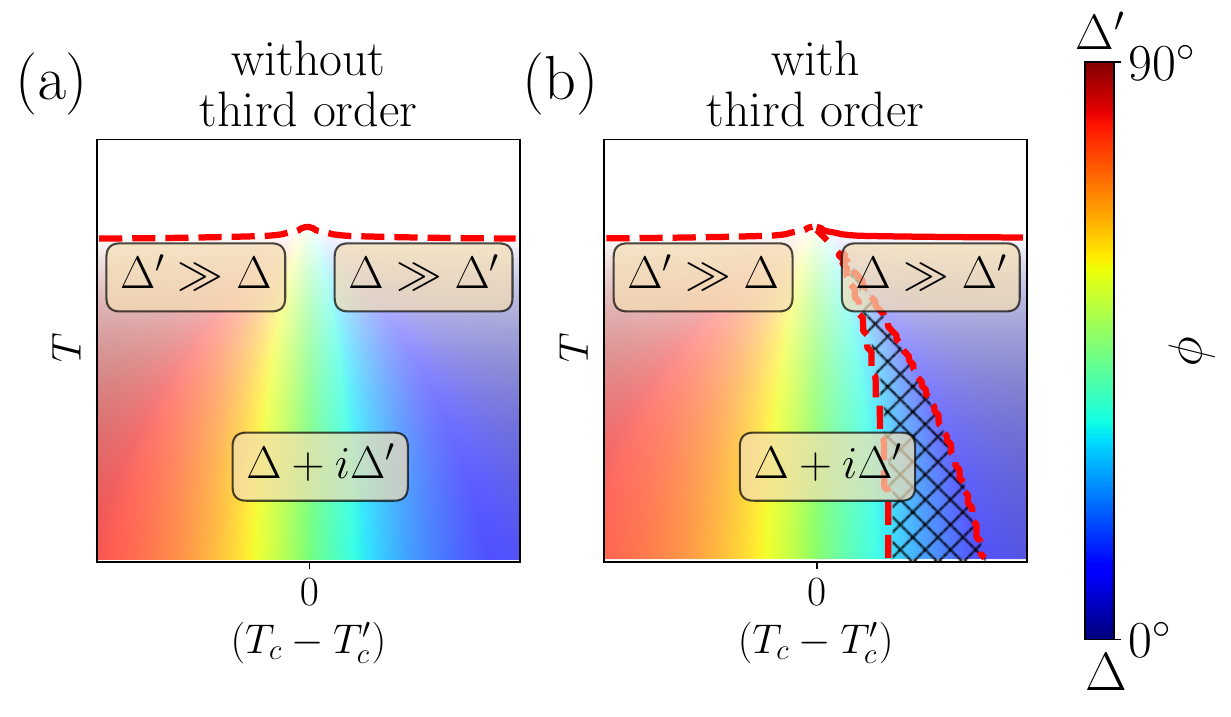}
    \caption{(a) Phase diagram in the presence of a magnetic field without a third order term applicable to the free energy $\mathcal{F}_{34}$. (b) Phase diagram in the presence of a magnetic field with a third order term applicable to the free energy $\mathcal{F}_{12}$. The color indicates the angle $\phi=\arctan \frac{\sum_i \Delta_i'^2}{\sum_i \Delta_i^2}$, while the intensity denotes $\sum_i (\Delta_i^2+\Delta_i'^2)$. Solid lines indicate first-order phase transitions and dashed lines indicate second-order transitions. Hatching denotes an anisotropic solution of the order parameter. TRS is broken everywhere due to the applied field. The Landau free energy coefficients are chosen to be $\alpha=\alpha'=\lambda_1=\lambda_2=\mu_i=\beta_1=-\beta_2=1$, $\lambda_{i>2}=0$.}
    \label{fig:phase diagrams}
\end{figure}

Figure ~\ref{fig:phase diagrams}b, finally, shows the phase diagram in the presence of a magnetic field, when there is a third-order term in the free energy. The only order parameter combination that has both the lowest-order coupling to a magnetic field and a third-order term is $F_1$ with $F_2'$, in other words the free energy $\mathcal{F}_{1 2}$. 
The two order parameters are again coupled and always appear together. The regions $\Delta\gg\Delta'$, $\Delta\sim\Delta'$, and $\Delta\ll\Delta'$ are separated by crossovers. Due to the third-order term, an anisotropy can be induced, if $\Delta$ and $\Delta'$ both become  large enough. In the region where $T_c>T_c'$, adding the magnetic field increases the strength of the flux order, thereby enhancing the anisotropy via the third-order term. This is shown explicitly in appendix \ref{sec:app_anis_B}. Again, TRS is trivially broken in the entire phase diagram.

\section{Constraints from experiments on AV$_3$S\MakeLowercase{b}$_5$}
\label{sec:exp_constr}
While our phenomenological description of charge density orders in kagome metals is valid for any such system, we comment in the following on the consequences of our discussion for the AV$_3$Sb$_5$ family.
There are several experimental facts, which any theory of the charge-ordered (normal) state should reproduce: 
\begin{enumerate}
    \item A $2\times2$ increase in the in-plane unit cell at $T_c$ \label{fact_2b2} as observed by X-ray diffraction~\cite{Ortiz2,Ortiz3,conjoined} and STM~\cite{Jiang,Zhao,Liang_STM,Chen}.
    \item The transition at $T_{\rm c}$ appears to be first order\label{fact_1o}. First, the heat capacity displays a sharp peak at $T_{\rm c}$~\cite{Ortiz2}, which is a general feature of a first-order transition. In addition, X-ray scattering~\cite{conjoined} and NMR~\cite{NMR} show discontinuities at $T_{\rm c}$, further suggesting a first-order transition. Finally, transport \cite{Ortiz,guo2023correlated} does not see an extended region of fluctuations but a rather abrupt change, especially for out-of-plane conductivity.
    \item TRS is/can be broken below $T'<T_{\rm c}$, \label{fact_trsb} as seen in muon spin-relaxation~\cite{Mielke,yu2021evidence} and Kerr rotation~\cite{Xu,hu2023timereversal}. It is currently uncertain whether this is truly spontaneous TRSB or whether it is a giant response to a small applied magnetic field. 
     In either case, it is clear that the coupling of the order to (out-of-plane) magnetic fields is important: A small field leads to a giant anomalous Hall response~\cite{Yu2,Yang} and  a large $\mu$SR response~\cite{Guguchia,Khasanov,Mielke}. Finally, such a magnetic field (linearly) couples the chirality of the ordered state~\cite{Guo}.
\end{enumerate}

Experimental fact \ref{fact_2b2} implies we should consider the translational-symmetry breaking orders of $C_{6v}'''$, in other words the four three-dimensional irreps denoted as $F_i$, $i=1,\ldots 4$. Experimental fact \ref{fact_1o} requires the presence of a third-order term in $\mathcal{F}$. Absent a third-order term, the phase-transitions in the Landau free energy are generically second order.  Experimental fact \ref{fact_trsb} suggests that the system is at least very tunable towards TRSB order, which should therefore lie close in energy to the ground state. This implies we should consider additional flux order, in other words $F_2'$ and $F_4'$. In addition, experimental fact \ref{fact_trsb} requires a term that (linearly) couples the magnetic field to the order parameters. In our classification, the only order-parameter combination that has third-order terms as well as (linear) coupling to a magnetic field is $F_1$ with $ F_2'$. There is evidence from experiments and density functional theory that the charge order transforms as the $F_1$ irrep \cite{Ratcliff,WithoutAcoustic,Blumberg,Yan,Uykur}, which supports the conclusion reached above using different means. The nature of the flux order has not been determined yet by other means.

\section{Proposals for future experiments on AV$_3$S\MakeLowercase{b}$_5$}
\label{sec:exp_proposals}

\subsection{Transport}
With the order parameter combination $F_1$ and $ F_2'$ we are able to explain the striking experimental results seen in transport in Ref.~\cite{guo2023correlated}. The experiment reported isotropic transport in the absence of strain, however application of an \emph{out-of-plane} magnetic field leads to anisotropic transport. If we are in the regime where $T_{\rm c}\gg T_{\rm c}'$, then only $\Delta$ is induced at $T_{\rm c}$ and this order is isotropic. However, a magnetic field will induce $\Delta'$ at $T_{\rm c}$ as well (due to the $\mu_1$ term in the free energy). The $\beta_2$ third-order term coupling $\Delta$ and $\Delta'$ can then lead to anisotropy as outlined in Sec.~\ref{sec:nem}.

One natural future direction to explore is how closely the transport anisotropy is related to the charge order. It has been shown that with sufficient Nb and Ta doping on the V-site, the chemical pressure can suppress the charge density wave transition down to lower or even zero temperature~\cite{YANG20222176,PhysRevB.105.L180507}. The report of an isotropic superconducting gap suggests the absence of anisotropy without applying an external field~\cite{Zhong2023}, but the field- or strain-induced anisotropy is yet to be explored. Therefore, we propose a complete mapping of anisotropy across the doping phase diagram in order to explore how the field- and strain- dependence of the electronic anisotropy evolves when the charge order is suppressed or absent. 

Meanwhile, previous experimental results showed that when a magnetic field is applied, the direction of transport anisotropy seems to be pinned to a particular direction most likely due to the small uniaxial strain~\cite{guo2023correlated}. It is worth checking whether this pinning can be altered by a strong current pulse which may overcome the barrier of the pinning energy. 

Moreover, the angular dependence of magnetoresistance (MR) seems to suggest a two-fold symmetry which indicates a breaking of the  six-fold in plane rotation symmetry~\cite{Xiang}. Yet, this technique is strongly limited by the misalignment of the magnetic field direction due to the huge magnetoresistance spike with in-plane field~\cite{Guo}. Therefore, measuring the angular dependence of the magnetoresistance with a spherical rotation of the magnetic field direction would reveal the true in-plane component of MR with and without an out-of-plane field component. 

Furthermore, we are working with the assumption that there is no anisotropy in the strain-free case. Elastoresistance measurements such as those in Ref.~\onlinecite{Nie} are interesting in order to quantify the dependence on strain. 

Most of the theoretical treatment of our work assumes a two-dimensional nature of the charge ordering. One important open question is how much of the physics of the charge density order is two-dimensional. In particular, a possible explanation of the data in Ref.~\onlinecite{guo2023correlated} observing isotropic transport is that while the transport in a given layer is anisotropic, the transport averages over different layers such that one obtains isotropic transport. In order to rule out this scenario, one could probe the transport in a two-dimensional film of the material. Indeed, there has been recent progress in manufacturing thin films of the kagome metals via exfoliation \cite{2d_1,2d_2,song2021competing}. It would be exciting to reach the monolayer limit and measure transport anisotropy in that case.

\subsection{STM}

  Another approach to detect whether the isotropic transport behaviour in the low strain devices of Ref.~\onlinecite{guo2023correlated} arises due to averaging over many domains/layers would be to perform local measurements of the anisotropy, for example via STM, on ultra-low strain devices. Reference~\onlinecite{guo2023correlated} reports anisotropic transport once a magnetic field is switched on. It would be interesting to perform an STM study (or indeed an X-ray scattering experiment) when the sample is in a magnetic field, in order to establish the source of this anisotropy. However, one should caution that STM is a surface probe and the physics probed at the surface may not necessarily be representative of the physics in the bulk.

\subsection{TRSB}

Currently the results on TRSB in Kagome metals are still controversial, in particular, there is no consensus on whether there is spontaneous TRSB in zero magnetic field or whether the TRSB is induced by the small training fields used in experiments. In terms of the probes used to investigate (spontaneous) TRSB in the Kagome metals, there have been both Kerr effect and muon spin rotation experiments. Another probe that could be used to detect the circulating currents predicted in these materials is neutron scattering. This technique has been successfully used to detect the loop currents in the pseudogap phase of the cuprates \cite{trsb_neutrons}.

Our Landau theory as well as recent experiments \cite{guo2023correlated} show that kagome metals can be very sensitive to strain. While these experiments showed that there was no observable dependence of the charge-ordering temperature on the strain, it has yet to be established whether the TRSB temperature depends on strain.

\subsection{Stiffness measurements}

Transport experiments have revealed that the anisotropy in strain-free samples increases when a magnetic field is applied and this should yield observable signatures in measurements of elements of the stiffness matrix, as we elaborate on below.

In addition to the coupling between the strain and the order parameters, there is also a contribution to the free energy coming from the elastic energy itself. Using the Voigt notation for the strain tensor, i.e.,~introducing the six-dimensional vector $\mathbf{\epsilon}=(\epsilon_{xx},\epsilon_{yy},\epsilon_{zz},2\epsilon_{yz},2\epsilon_{xz},2\epsilon_{xy})$, we can write the elastic energy as 
\begin{equation}
    \mathcal{F}^{(\epsilon)}=\frac{1}{2}\sum_{\alpha\beta}c_{\alpha\beta}\epsilon_\alpha\epsilon_\beta,
\end{equation}
where $c_{\alpha\beta}$ is the stiffness matrix. For a system with $D_{6h}$ symmetry, there are five independent components of the stiffness matrix \cite{Grandi}: $c_{11},c_{12},c_{13},c_{33},c_{44}$. In general, the components of the stiffness matrix can be discontinuous at a second-order phase transition, which can be measured in experiments. The discontinuities are given by the formula \cite{Grandi,RUS_SRO,strain}
\begin{equation}
    \Delta c_{\alpha\beta}=\sum_{\gamma\delta}\frac{\partial^2\mathcal{F}^\textrm{(str)}}{\partial\epsilon_\alpha\partial D_\gamma}\frac{\partial^2\mathcal{F}^\textrm{(str)}}{\partial\epsilon_\beta\partial D_\delta}\bigg(\frac{\partial^2\mathcal{F}}{\partial D_\gamma\partial D_\delta}\bigg)^{-1},
\end{equation}
where $\vec D=(\vec \Delta,\vec \Delta')$ is the six-dimensional vector combining the order parameters. As shown in Ref.~\onlinecite{Grandi}, at the onset of isotropic charge ordering, there will be discontinuities $\Delta c_{11}, \Delta c_{12},\Delta c_{13},\Delta c_{33}$. On the other hand, at the onset of anisotropic order, there will be further independent components of the stiffness matrix, which are discontinuous, namely $\Delta c_{22}\neq\Delta c_{11}$ and $\Delta c_{13}\neq\Delta c_{23}$. Therefore, one interesting experiment would be to measure $\Delta c_{22}-\Delta c_{11}$ and $\Delta c_{13}-\Delta c_{23}$ in a pristine (strain-free) sample as a function of an applied magnetic field. 

Resonant ultrasound spectroscopy is a method used to measure the elements of the elastic stiffness matrix~\cite{RUS}. It would be interesting to perform these experiments as a function of the applied magnetic field, though resonant ultrasound spectroscopy in a magnetic field is challenging and it is possible that pulse-echo sound velocity measurements are more realistic.

\section{Conclusion}
\label{sec:concl}
Kagome metals are known to undergo charge ordering with a $2\times 2$ increase in the size of the unit cell. Using a group theory analysis, we write down all possible site, bond and flux ordering that is consistent with the system symmetries. The observation of TRSB and coupling to a magnetic field further motivates studying TRSB flux order. Flux and bond order are natural partners since they are the imaginary and real part of the same underlying order parameter and so in general, we expect these two orders to be coupled. This leads us to study \emph{all} possibilities for the coupling of flux and bond order. 

We use a Landau analysis to study the interplay of the two orders. The different types of flux and bond order lead to differences in the third order term in the Landau theory as well as differences in the coupling to the magnetic field. We construct the phase diagrams for different types of coupled flux and bond order. Depending on the type of order, the two order parameters corresponding to flux and bond order may appear separately or always in unison. 

By synthesizing various experimental results and comparing to the Landau theory phase diagrams, we are able to deduce that the most likely candidate is a tri-hexagonal or Star of David bond order with a subsidiary $C_2$-preserving flux order. 
\\
\\
\textbf{Note added:} After completion of our work, a new experimental preprint appeared \cite{xing2023optical}, which also supports the irrep combination $F_1$ with $F_2'$ for the in-plane ordering in the material RbV$_3$Sb$_5$.
\\
\\
\section{Acknowledgements}
We thank F.~Grandi and M.~Christensen for comments on a previous version of the manuscript. We also thank R. Fernandes for helpful discussions. \textcolor{black}{We thank Sofie Castro Holbaek for noticing an error in a previous version of this manuscript.} This project was
supported by the European Research Council (ERC) under the European Union’s Horizon 2020 research and innovation program grant nos.~ERC-StG-Neupert-757867-PARATOP (GW, TN) and 715730 (CG, PJWM). TN acknowledges funding from the Swiss National Science Foundation (Project 200021E 198011) as part of the QUAST FOR 5249-449872909 (Project P3).

 	\bibliographystyle{unsrtnat}
 	\bibliography{bib.bib}

\clearpage 
\newpage

\onecolumngrid
	\begin{center}
		\textbf{\large --- Supplementary Material ---\\ Phenomenology of bond and flux orders in kagome metals}\\
		\medskip
		\text{Glenn Wagner, Chunyu Guo, Philip J. W. Moll, Titus Neupert, Mark H. Fischer}
	\end{center}
	
		\setcounter{equation}{0}
	\setcounter{figure}{0}
	\setcounter{table}{0}
	\setcounter{page}{1}
	\makeatletter
 \renewcommand{\theequation}{\thesection\arabic{equation}}
	\renewcommand{\thefigure}{S\arabic{figure}}
	\renewcommand{\bibnumfmt}[1]{[S#1]}
\begin{appendix}

\section{Group theory details}
\label{app:decomp}

\begin{table}[h]
\begin{tabular}{|l|l|l|l|l|l|l|l|l|l|l|}
\hline
                               & $I$ & $t_i$ & $C_2$ & $t_iC_2$ & $C_3$ & $C_6$ & $\sigma_v$ & $t_i\sigma_v$ & $\sigma_d$ & $t_i\sigma_d$ \\ \hline
$|\mathcal{C}|$ & 1   & 3     & 1     & 3        & 8     & 8     & 6          & 6             & 6          & 6             \\ \hline
$A_1$                          & 1   & 1     & 1     & 1        & 1     & 1     & 1          & 1             & 1          & 1             \\
$A_2$                          & 1   & 1     & 1     & 1        & 1     & 1     & -1         & -1            & -1         & -1            \\
$B_1$                          & 1   & 1     & -1    & -1       & 1     & -1    & 1          & 1             & -1         & -1            \\
$B_2$                          & 1   & 1     & -1    & -1       & 1     & -1    & -1         & -1            & 1          & 1             \\
$E_1$                          & 2   & 2     & -2    & -2       & -1    & 1     & 0          & 0             & 0          & 0             \\
$E_2$                          & 2   & 2     & 2     & 2        & -1    & -1    & 0          & 0             & 0          & 0             \\\hline
$F_1$                          & 3   & -1    & 3     & -1       & 0     & 0     & 1          & -1            & 1          & -1            \\
$F_2$                          & 3   & -1    & 3     & -1       & 0     & 0     & -1         & 1             & -1         & 1             \\
$F_3$                          & 3   & -1    & -3    & 1        & 0     & 0     & 1          & -1            & -1         & 1             \\ 
$F_4$                          & 3   & -1    & -3    & 1        & 0     & 0     & -1         & 1             & 1          & -1            \\ \hline
\end{tabular}
\caption{Character table of the group $C_{6v}'''$.}
\label{tab:character_table_app}
\end{table}

For the site order, we compute the permutation matrix $\mathcal{P}_s(g)$ that describes how the sites transform into each other under the operation $g\in C_{6v}'''$. Since there are 12 sites in the $2\times2$ enlarged unit cell, the set of permutation matrices describe a 12-dimensional reducible representation of $C_{6v}'''$. In order to decompose this representation into irreps, we compute the characters $\chi(g)=\textrm{Tr}(\mathcal{P}_s(g))$, in other words, we count the number of sites that map to themselves under the operations of $C_{6v}'''$. We find
 \begin{table}[h!]
 \centering
 \begin{tabular}{p{2.5cm} p{1.3cm} p{1.3cm} p{1.3cm} p{1.3cm} p{1.3cm} p{1.3cm} p{1.3cm} p{1.3cm} p{1.3cm} p{1.3cm} p{1.3cm}}  
$g$ & $I$ & $t_i$ & $C_2$ & $t_iC_2$ & $C_3$ & $C_6$ & $\sigma_v$ & $t_i\sigma_v$ & $\sigma_d$ & $t_i\sigma_d$\\ \hline
 $\textrm{Tr}(\mathcal{P}_s(g))$ & 12        & 0         & 0         & 4        & 0    & 0       & 2 &0 & 2 & 0  
 \label{tab:coefs}
 \end{tabular}
 \end{table}
\\
We can then compute the multiplicity $n_R$ of the irrep $R$ in the decomposition of this reducible represention via
\begin{equation}
    n_R=\frac{1}{|C_{6v}'''|}\sum_{g\in C_{6v}'''}\chi_R(g)\chi(g),
\end{equation}
where the characters $\chi_R(g)$ are listed in the character table (Tab.~\ref{tab:character_table_app}). This leads to the decomposition of site order
\begin{equation}
    \mathcal{P}_s=A_1+E_2+F_1+F_3+F_4.
\end{equation}
The bond and flux order can be treated similarly.

\section{Multiplication table of irreps of $C_{6v}'''$}
We want to decompose the product of two irreps $R_i$ and $R_j$ into a sum of irreps \begin{equation}
    R_i\otimes R_j=\bigoplus_kn_kR_k,
\end{equation}
where 
\begin{equation}
    n_k=\frac{1}{|C_{6v}'''|}\sum_{g\in C_{6v}'''}\chi_k(g)\chi_i(g)\chi_j(g). 
\end{equation}
We list the results in Tab.~\ref{tab:MT}.

\section{Coupling to strain}
\label{app:strain}
The crucial symmetry to determine the coupling to strain is $C_3$. The components of the strain tensor transform in the $E$ irrep, which transforms non-trivially under $C_3$, and so in order to obtain a scalar (i.e.~a term that we can add to the free energy), we need to construct a term out of the order parameter that also transforms under the $E$ irrep. The basis functions of the $E$ irrep are conventionally labelled as $p_\pm=p_x\pm ip_y$ which pick up phases of $\omega=e^{2\pi i/3}$ and $\omega^*$ respectively under $C_3$. Looking at the transformation properties of the order parameter under $C_3$, we find that the quadratic terms that transform correctly are 
\begin{align}
    p_+&=\Delta_1^2+\omega \Delta_2^2+\omega^2\Delta_3^2,\\
    p_-&=\Delta_1^2+\omega^2 \Delta_2^2+\omega \Delta_3^2,
\end{align}
where $\omega=e^{2\pi i/3}$. Then we write these as $p_\pm=(p_x\pm ip_y)$ with
\begin{align}
    p_x&= \Delta_1^2- \Delta_2^2/2- \Delta_3^2/2,\\
    p_y&=\frac{\sqrt{3}}{2}(\Delta_2^2- \Delta_3^2).
\end{align}
One can check that the doublet $\{p_x,p_y\}$ transforms in the same way under $C_3$ as $\{\epsilon_{p_x},\epsilon_{p_y}\}=\{(\epsilon_{xx}-\epsilon_{yy})/2 ,\epsilon_{xy}\}$, allowing us to construct the $C_3$-symmetric term to be added to the free energy:
\begin{equation}
    \mathcal{F}^{(\textrm{str})}=2\mu_2(\epsilon_{p_x}p_x+\epsilon_{p_y}p_y)
    =\mu_2[(\epsilon_{xx}-\epsilon_{yy})(\Delta_1^2-\Delta_2^2/2-\Delta_3^2/2)+\epsilon_{xy}\sqrt{3}(\Delta_2^2-\Delta_3^2)].
\end{equation}
The analogous term for flux order automatically respects time-reversal symmetry since it is quadratic in the order parameter and hence flux order couples to strain in the same manner (with a different coupling coefficient). 

The coupling to strain gives us a way to quantify the anisotropy in the system. For an isotropic phase, the order parameters $p_x$ and $p_y$ will be zero. If the solution is anisotropic, there will be degenerate solutions with different values of $p_x$ and $p_y$ and application of strain will pick out one of these degenerate solutions leading to divergent susceptibility. Let us consider the response to strain in the system at finite temperature $T$. The expectation value of the symmetry-breaking order parameter is
\begin{equation}
    \langle p_x\rangle_{\mathcal{F}(\epsilon_{p_x}),T}=\langle p_xe^{-\beta\mathcal{F}^{(\textrm{str})}}\rangle_{0,T}=\langle p_xe^{-2\beta\mu_2\epsilon_{p_x}p_x}\rangle_{0,T},
\end{equation}
where $\langle\dots\rangle_{0,T}$ denotes the finite-temperature average with respect to the zero-strain free energy $\mathcal{F}(\epsilon_{p_x}=0)$ and $\beta=1/(k_BT)$.
The susceptibility is then
\begin{equation}
    \chi_{p_x}=\lim_{\epsilon_{p_x}\to0}\frac{\partial \langle p_x\rangle_{\mathcal{F}(\epsilon_{p_x}),T}}{\partial \epsilon_{p_x}}=-2\beta\mu_2\langle p_x^2\rangle_{0,T},
\end{equation}
and similarly $\chi_{p_y}=-2\beta\mu_2\langle p_y^2\rangle_{0,T}$. The susceptibilities diverge when $T\to0$ if $p_x^2$ or $p_y^2$ acquire a finite expectation value in the ground state. This is the signature of spontaneous symmetry breaking. This motivates us to introduce the anisotropic order parameter
\begin{equation}
    \frac{1}{\beta}\textrm{Tr}\chi= \frac{1}{\beta}(\chi_{p_x}+\chi_{p_y})=-2\mu_2(p_x^2+p_y^2)=-4\beta\mu_2\sum_{i,j}(\Delta_i^2-\Delta_j^2)^2
\end{equation}
and an analogous order parameter can be written down for the flux order.

\section{First-order transition}
\label{app:pd}

\begin{figure}[h!]
  \subfloat[case with $d=0$]{
      \includegraphics[width=0.5\textwidth]{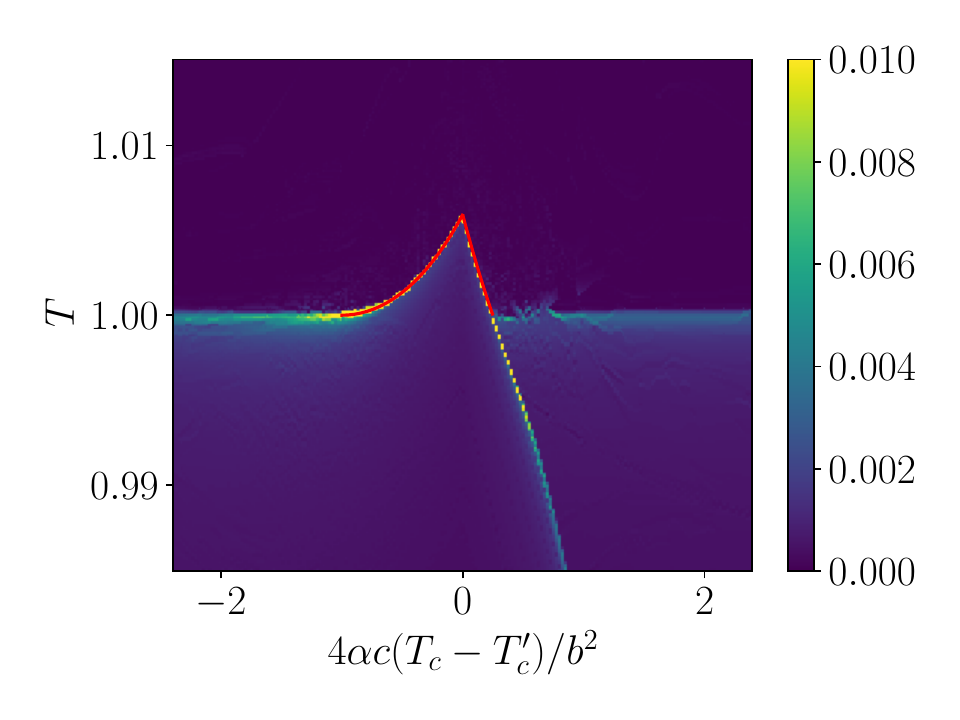}}\hfill
  \subfloat[case with $d=b/4$]{
      \includegraphics[width=0.5\textwidth]{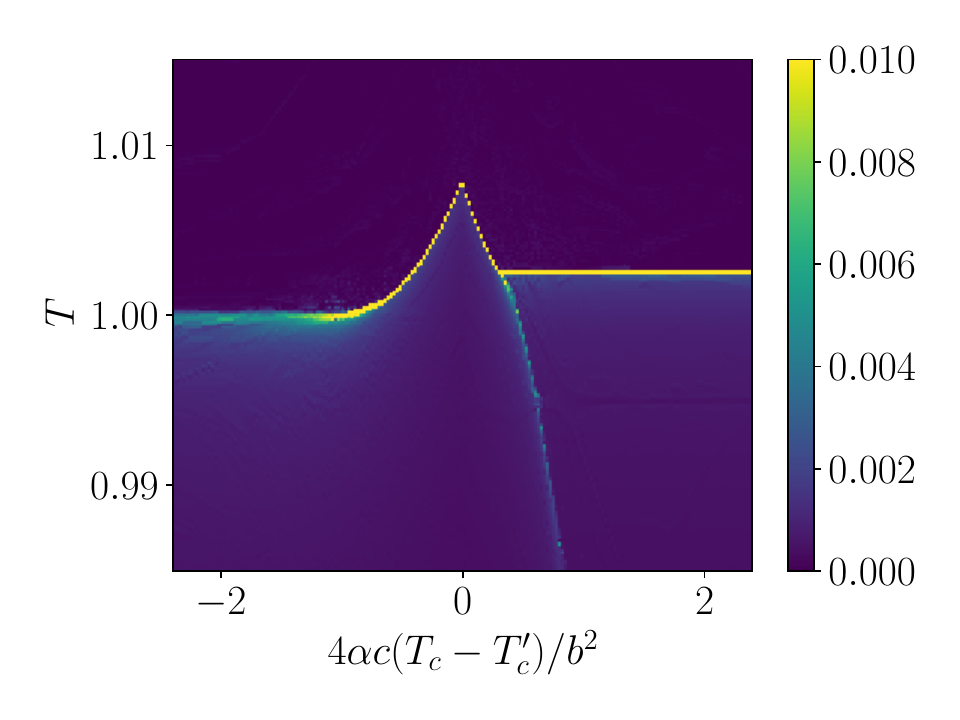}}\hfill
  \caption{Temperature derivative of the order parameter amplitude $\Delta^2+\Delta'^2$ for the free energy \eqref{eq:Free_energy} with a term $d\Delta^3$. Large values of the derivative (yellow) indicate first order phase transitions. (a) In the case $d=0$, there is a first order transition for a finite range of $T_c-T_c'$. The red curves show the analytical results \eqref{eq:ana1} and \eqref{eq:ana2}. (b) In the case $d>0$, there is a first order phase transition for all $T_c>T_c'$.}
  \label{fig:3rd_order}
\end{figure}

When $T_c\sim T_c'$, there is a first-order transition into a phase with both charge and flux order. Let us consider the free energy
\begin{align}
    \mathcal{F}&=\alpha(T-T_c)\Delta^2+\alpha(T-T_c')\Delta'^2+b\Delta\Delta'^2+c(\Delta^2+\Delta'^2)^2=A\psi^2+B\psi^3+C\psi^4
    \label{eq:Free_energy}
\end{align}
where we used the parametrization $\Delta+i\Delta'=\psi e^{i\phi}$ and
\begin{align}
    A=&\alpha(T-T_c)\cos^2\phi+\alpha(T-T_c')\sin^2\phi=\alpha[T-T_c\cos^2\phi-T_c'\sin^2\phi]\\
    B=&b\cos\phi\sin^2\phi\\
    C=&c.
\end{align}
The minimum of the free energy is at a non-zero value of the order parameter when $A<\frac{B^2}{4C}$ i.e.~when
\begin{equation}
    T<T_1=T_c\cos^2\phi+T_c'\sin^2\phi+\frac{b^2}{4\alpha c}\cos^2\phi\sin^4\phi=T_c+(T_c'-T_c)[\sin^2\phi+\eta(\sin^4\phi-\sin^6\phi)],
\end{equation}
where $\eta=b^2/[4\alpha c(T_c'-T_c)]$. We will have a first-order transition, when there is a non-zero solution at $T>T_c,T_c'$. Consider first the case where $T_c'>T_c$. Maximizing the term in square brackets leads to 
\begin{equation}
    \text{max}_\phi(T_1)=T_c+(T_c'-T_c)\frac{(\eta+\sqrt{\eta(3+\eta)})(6+\eta+\sqrt{\eta(3+\eta)})}{27\eta},
    \label{eq:ana1}
\end{equation}
when $\sin^2\phi=(\eta+\sqrt{\eta(3+\eta)})/(3\eta)\leq1$. We have $\text{max}_\phi(T_1)>T_c'$ when $\eta>1$. Similarly, when $T_c>T_c'$, we can write
\begin{equation}
    T_1=T_c'+(T_c-T_c')[\cos^2\phi-\eta\cos^2\phi(1-\cos^2\phi)^2].
\end{equation}
Maximizing with respect to $\phi$ yields
\begin{equation}
    \text{max}_\phi(T_1)=T_c'+(T_c-T_c')\frac{2(2\eta+\sqrt{\eta(3+\eta)})(3-\eta+\sqrt{\eta(3+\eta)})}{27\eta},
    \label{eq:ana2}
\end{equation}
and $\text{max}_\phi(T_1)>T_c$ when $\eta<-4$. So we have a first-order transition in a range where the critical temperatures of the two orders are similar:
\begin{equation}
    \frac{b^2}{4\alpha c}>(T_c'-T_c)>-\frac{b^2}{16\alpha c}.
\end{equation}

\section{Anisotropy from third-order term}
\label{sec:app_anis}

\begin{figure}[h]
    \centering
    \includegraphics[width=\textwidth]{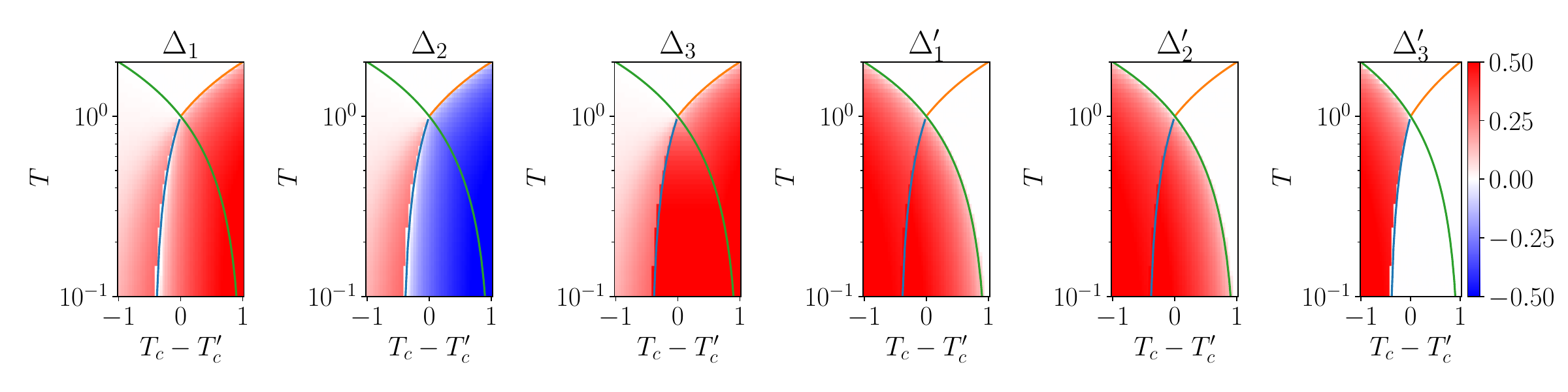}
    \caption{We plot the six order parameters and the phase boundaries between the different solutions. The blue line shows the boundary between solution 1 and solution 2 ($\sim T^*$), while the green line shows the boundary between solution 3 and solution 4 ($\sim T_c'$). In between the blue and green lines there is a crossover from solution 2 to solution 3. The yellow lines shows the onset of $\Delta$ order ($\sim T_c$). The wedge between the blue and green lines is where the solution is anisotropic.}
    \label{fig:anisotropy}
\end{figure}

Anisotropy can arise from the third-order term in the free energy even when the fourth-order terms favor an isotropic solution. To see this, consider the terms
\begin{equation}
\mathcal{F}^{(3)}=\beta_1\Delta_1\Delta_2\Delta_3+\beta_2(\Delta_1\Delta_2'\Delta_3'+\Delta_1'\Delta_2\Delta_3'+\Delta_1'\Delta_2'\Delta_3)
\end{equation}
as perturbations with $\Delta^2=\Delta_1^2+\Delta_2^2+\Delta_3^2$ and $\Delta'^2=\Delta_1'^2+\Delta_2'^2+\Delta_3'^2$ being fixed by the second-order and fourth-order terms in the free energy. The results derived below will therefore be exact in the limit $\beta_1\to0,\ \beta_2\to0$, though the numerics demonstrate that the results are similar for finite $\beta_1,\beta_2$. We study the case $\beta_1\beta_2<0$. The signs of $\beta_1$ and $\beta_2$ can always be flipped by changing the sign of $\Delta_i$, so without loss of generality, we assume $\beta_1>0$ and $\beta_2<0$. We now study the competition between four different solutions of the free energy, which will be the ground states as we traverse the phase diagram in Fig.~\ref{fig:anisotropy} from left to right. 
\begin{itemize}
    \item Solution 1: $\Delta_1=\Delta_2=\Delta_3=\frac{\Delta}{\sqrt{3}}$ and $\Delta_1'=\Delta_2'=\Delta_3'=\frac{\Delta'}{\sqrt{3}}$ leads to $\mathcal{F}^{(3)}=\frac{|\beta_1|}{3\sqrt{3}}\Delta^3-\frac{|\beta_2|}{\sqrt{3}}\Delta\Delta'^2$
    \item Solution 2: $\Delta_3=\Delta$ and $\Delta_1'=\Delta_2'=\frac{\Delta'}{\sqrt{2}}$ leads to $\mathcal{F}^{(3)}=-\frac{|\beta_2|}{2}\Delta\Delta'^2$
    \item Solution 3: $\Delta_1=-\Delta_2=\Delta_3=\frac{\Delta}{\sqrt{3}}$ and $\Delta_1'=\Delta_2'=\frac{\Delta'}{\sqrt{2}}$ leads to $\mathcal{F}^{(3)}=-\frac{|\beta_1|}{3\sqrt{3}}\Delta^3-\frac{|\beta_2|}{2\sqrt{3}}\Delta\Delta'^2$
    \item Solution 4: $\Delta_1=-\Delta_2=\Delta_3=\frac{\Delta}{\sqrt{3}}$ and $\Delta_i'=0$ leads to $\mathcal{F}^{(3)}=-\frac{|\beta_1|}{3\sqrt{3}}\Delta^3$
\end{itemize}
Let us further assume second-order terms $\mathcal{F}^{(2)}=\alpha(T-T_c)\Delta^2+\alpha(T-T_c')\Delta'^2$ and fourth-order terms $\mathcal{F}^{(4)}=\lambda(\Delta^4+\Delta'^4)$ such that
\begin{equation}
    \Delta^2=\frac{\alpha(T_c-T)}{2\lambda},
\end{equation}
\begin{equation}
    \Delta'^2=\frac{\alpha(T_c'-T)}{2\lambda}.
\end{equation}
Comparing the energies of solution 1 and solution 2, we find that the transition occurs at
\begin{equation}
    T^*=\frac{T_c-\zeta T_c'}{1-\zeta},
    \label{eq:Ts}
\end{equation}
where $\zeta=3(1-\frac{\sqrt{3}}{2})\vert\beta_2\vert/\vert\beta_1\vert$. We plot this transition temperature as a blue line in Fig.~\ref{fig:anisotropy}. Solutions 1 and 4 are isotropic, while solutions 2 and 3 are anisotropic. 

\section{Anisotropy in the presence of a magnetic field}
\label{sec:app_anis_B}

\begin{figure}[h]
    \centering
    \includegraphics[width=\textwidth]{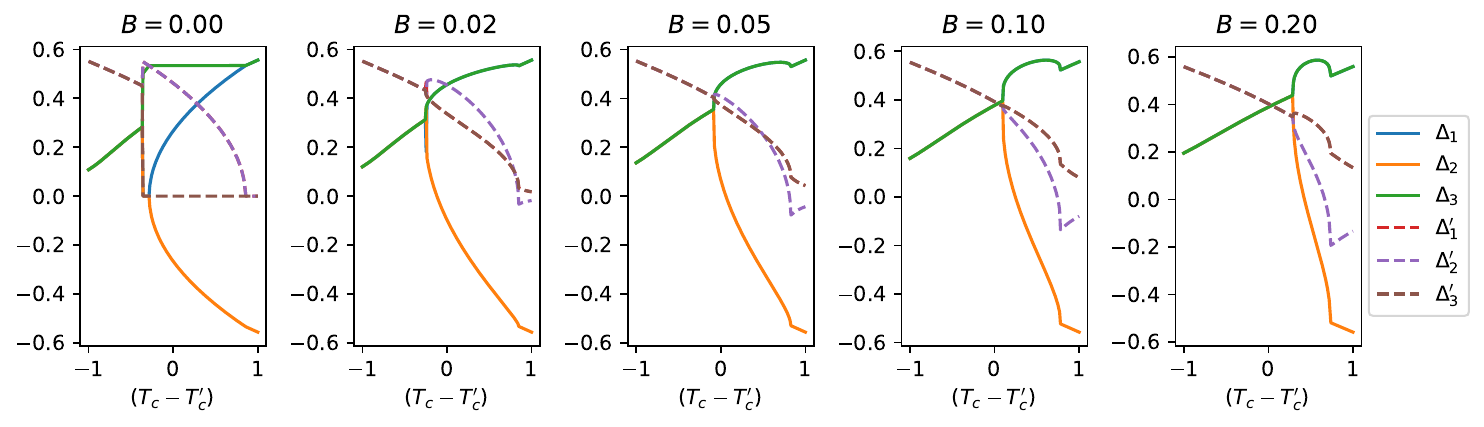}
    \caption{We plot the six order parameters for a set of different magnetic fields. The order parameters are computed at a fixed temperature as a function of $T_c-T_c'$. For $B=0$ we see the transitions between the four different solutions described in appendix \ref{sec:app_anis}. The region where solution is most stable (at the left of the phase diagram) becomes larger as the $B$-field increases.}
    \label{fig:Bdep}
\end{figure}

\begin{figure}[h]
    \centering
    \includegraphics[width=0.5\textwidth]{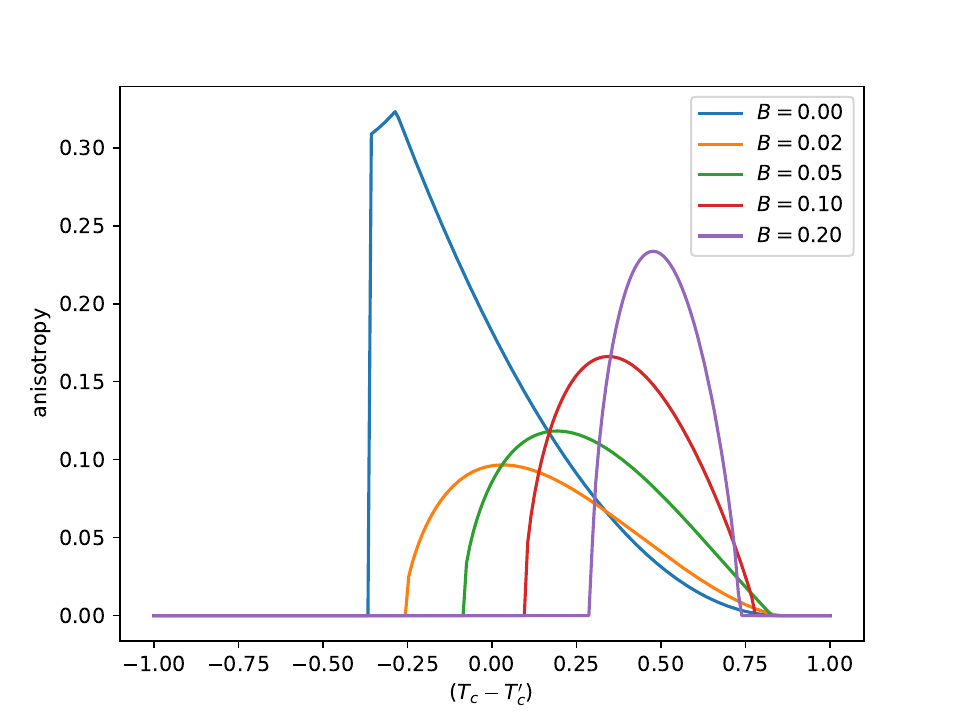}
    \caption{We plot order parameter of the anisotropy $\sum_{i,j}[(\Delta_i^2-\Delta_j^2)^2+(\Delta_i^{\prime2}-\Delta_j^{\prime2})^2]$ for a set of different magnetic fields. We fix the temperature $T=0.2$. We plot the anisotropy at a fixed temperature as a function of $T_c-T_c'$. There is a range of $T_c-T_c'$, where the magnetic field significantly enhances the anisotropy.}
    \label{fig:app_anisotropy}
\end{figure}

\begin{figure}[h]
    \centering
    \includegraphics[width=0.5\textwidth]{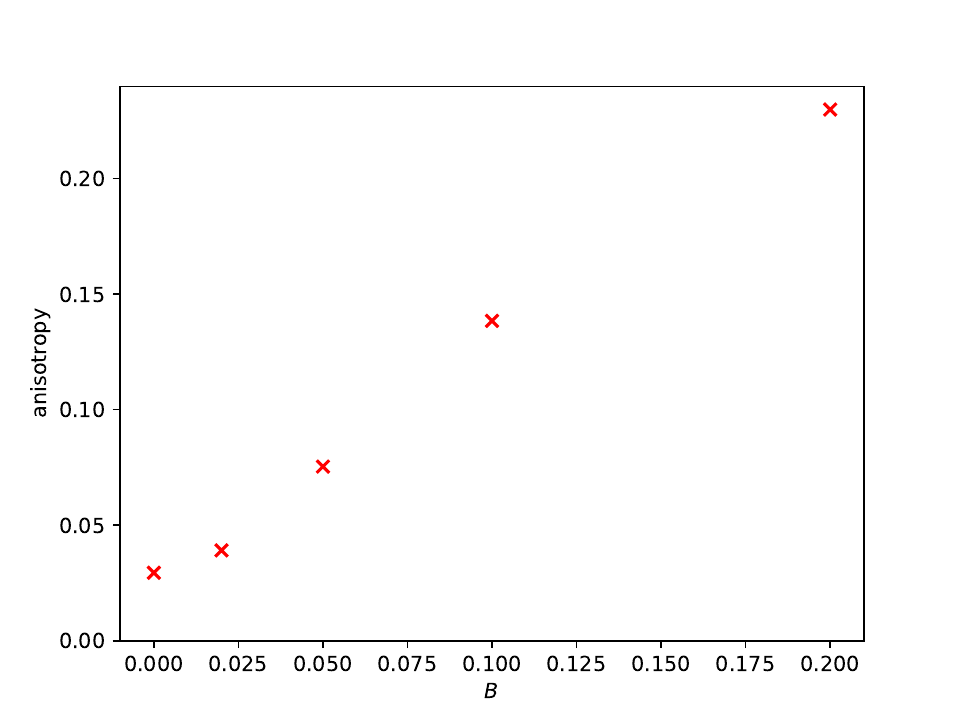}
    \caption{Increase of the anisotropy $\sum_{i,j}[(\Delta_i^2-\Delta_j^2)^2+(\Delta_i^{\prime2}-\Delta_j^{\prime2})^2]$ as a function of the applied field. We fix $T_c-T_c'=0.5$ and $T=0.2$. }
    \label{fig:B_anis_0.5}
\end{figure}

\begin{figure}[h!]
  \subfloat[case with $T_c-T_c'=-0.5$]{
      \includegraphics[width=0.5\textwidth]{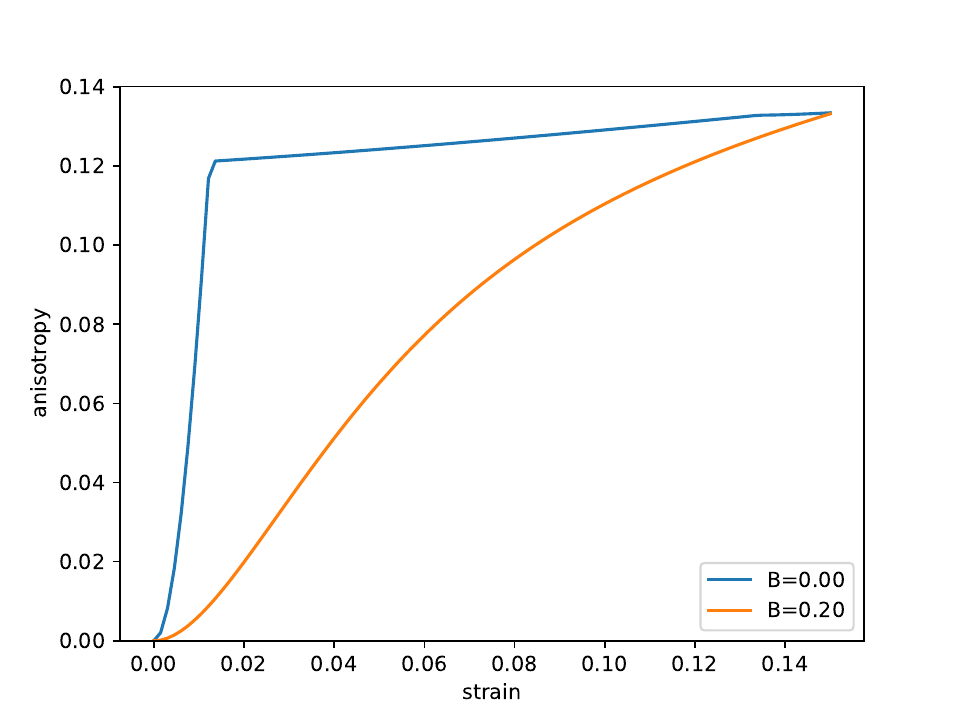}}\hfill
  \subfloat[case with $T_c-T_c'=0.5$]{
      \includegraphics[width=0.5\textwidth]{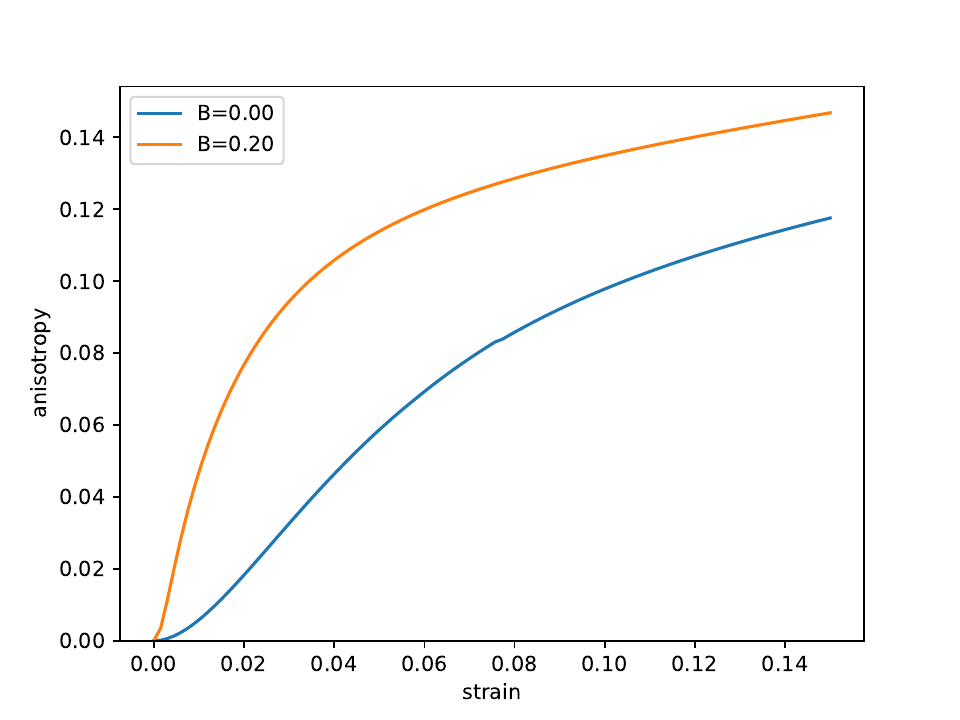}}\hfill
  \caption{Increase of the anisotropy $\sum_{i,j}[(\Delta_i^2-\Delta_j^2)^2+(\Delta_i^{\prime2}-\Delta_j^{\prime2})^2]$ as a function of the applied strain for two different values of the magnetic field. We fix $T=0.55$. In the case $T_c>T_c'$ the application of the magnetic field increases the anisotropy, consistent with the experimental results from Ref.~\cite{guo2023correlated}.}
  \label{fig:strain_dep}
\end{figure}

The magnetic field adds a term to the free energy 
\begin{equation}
    \mathcal{F}^{(B)}=\mu_1B(\Delta_1\Delta_1'+\Delta_2\Delta_2'+\Delta_3\Delta_3').
\end{equation}
The effect of the magnetic field is to induce $\Delta'$ as soon as $\Delta$ is present. Let us assume without loss of generality $B<0$ (we can always switch the sign of $\Delta_i'$ to adapt to the sign of $B$). Of the four solutions, the only solution that can take advantage of $\mathcal{F}^{(B)}$ and lower its energy is solution 1, which has a contribution $-B\Delta\Delta'$. None of the other solutions have changes in energy to this order in the perturbation theory. Therefore solution 1 becomes more stable and the boundary between solution 1 and solution 2 shifts to the right, this is shown in Fig.~\ref{fig:Bdep}.  The result of this effects is that the magnetic field can increase the anisotropy of the order parameters, as shown in Figs.~\ref{fig:app_anisotropy} and \ref{fig:B_anis_0.5}. In particular, due to the phase boundary between solution 1 and solution 2 shifting to the right, the value of $\Delta$ is larger, when the components become anisotropic and this results in a larger anisotropy. We note that if we are in the part of the phase diagram where $T_c<T_c'$ then the magnetic field has the opposite effect and suppresses the anisotropy (Fig.~\ref{fig:strain_dep}).

\end{appendix}
\end{document}